\documentclass[review,hidelinks,onefignum,onetabnum]{siamart250211}

\nolinenumbers

\usepackage{lipsum}
\usepackage{amsfonts}
\usepackage{graphicx}
\usepackage{epstopdf}
\usepackage{algorithmic}
\usepackage{subfigure}
\usepackage{multirow}
\ifpdf
  \DeclareGraphicsExtensions{.eps,.pdf,.png,.jpg}
\else
  \DeclareGraphicsExtensions{.eps}
\fi

\newsiamremark{remark}{Remark}
\newsiamremark{hypothesis}{Hypothesis}
\crefname{hypothesis}{Hypothesis}{Hypotheses}
\newsiamthm{claim}{Claim}
\newsiamremark{fact}{Fact}
\crefname{fact}{Fact}{Facts}

\headers{AI prediction of ground states in rotating BEC}{Z. Kong, J. Z. Yang, C. Yuan, and X. Zhao}

\title{Toward fast, accurate and robust AI prediction of ground states in rotating BEC\thanks{Submitted to the editors DATE.
\funding{J. Z. Yang is supported by NSFC (No.12125103, No.U24A2002) and Hubei Natural Science Foundation (No.2025AFA002); X. Zhao is supported by National Key Research and Development Program of China (No. 2024YFE03240400), National MCF Energy R\&D Program and
NSFC 42450275, 12271413; C. Yuan is supported by NSFC (No.12301558).}}}

\author{Zhizhong Kong\thanks{School of Mathematics and Statistics, Wuhan University (\email{zhizhongkong@whu.edu.cn}).}
\and Jerry Zhijian Yang\thanks{School of Mathematics and Statistics \& Computational Sciences Hubei Key Laboratory, Wuhan University (\email{zjyang.math@whu.edu.cn}).}
\and Cheng Yuan\thanks{School of Artificial Intelligence, Wuhan University (\email{yuancheng@whu.edu.cn}).}
\and Xiaofei Zhao\thanks{School of Mathematics and Statistics \& Computational Sciences Hubei Key Laboratory, Wuhan University (\email{matzhxf@whu.edu.cn}).}}

\usepackage{amsopn}

\ifpdf
\hypersetup{
  pdftitle={Toward fast, accurate and robust AI prediction of ground states in rotating BEC},
  pdfauthor={Z. Kong, J. Z. Yang, C. Yuan, and X. Zhao}
}
\fi

\begin{document}

\maketitle

\begin{abstract}
We propose an unsupervised deep learning approach for computing the ground state (GS) of rotating Bose-Einstein condensation. To minimize the energy under a mass constraint, our approach introduces two key and novel ingredients: a normalized loss function that exactly enforces the mass constraint, and a training strategy named virtual rotation acceleration that is essential for avoiding local minima and guiding the learning process to the correct quantized vortex phase. Extensive numerical experiments demonstrate the proposed approach as an effective and accurate method to predict GS across physical conditions--from slow to fast rotation and from isotropic to anisotropic confinement. Through further distillation, we establish a unified operator network capable of efficiently generalizing physical parameters across different phases. It enables rapid GS predictions while correctly capturing phase transitions and is applied for inverse problems.
\end{abstract}



\section{Introduction}
The concept of Bose-Einstein condensation (BEC) emerged in 1924–1925, stemming from A. Einstein's extension of S.N. Bose's quantum statistics for photons to Bose gases \cite{einstein2005quantentheorie}. 
For a dilute Bose gas below the critical temperature and confined in an external trap with possible rotation, the mean-field approximation \cite{Lieb} employed for modeling leads to the following rotating Gross-Pitaevskii equation (GPE) in dimensionless form \cite{bao2006efficient,fetter2009rotating,BaoCai}:
\begin{equation}\label{model}
    i\frac{\partial\Phi(\mathbf{x},t)}{\partial t} = -\frac{1}{2}\nabla^{2}\Phi(\mathbf{x},t) + V(\mathbf{x})\Phi(\mathbf{x},t) + \beta|\Phi(\mathbf{x},t)|^{2}\Phi(\mathbf{x},t)-\Omega L_z\Phi(\mathbf{x},t),
\end{equation}
where $t$ is the time, $\mathbf{x} = (x,y)^{\top} \in \mathbb{R}^{2}$ or $\mathbf{x} = (x,y,z)^{\top} \in \mathbb{R}^{3}$ is the spatial variable, $\Phi(\mathbf{x},t)$ is the unknown complex-valued wave function and $L_z = -i(x\partial_y-y\partial_x)$ is the angular momenta operator with $\Omega\geq0$ the rotating speed. The parameter $\beta\in\mathbb{R}$ is a given constant with $\beta>0$ denoting the repulsive self-interaction and $\beta<0$ denoting the attractive case, and $V(\mathbf{x})$ is a given real-valued potential function satisfying
$\lim_{|\mathbf{x}|\to\infty} V(\mathbf{x})=\infty$ for confinement.

The first BEC was realized in experiments in 1995 \cite{1stBEC} and later quantized vortices were discovered in the rotating frame \cite{vortexexp,fastvortex,fetter2009rotating,jackson1998vortex}, revealing the superfluid property of BEC and also promoting GPE \eqref{model} for wide applications that ranges from atoms to cosmology. In (\ref{model}), the total mass $M$ and energy $E$ are conserved, i.e.,
$$
M(\Phi(\cdot, t)) := \int_{\mathbb{R}^{d}} |\Phi(\mathbf{x}, t)|^{2} \, d\mathbf{x} \equiv M(\Phi(\cdot, 0)), \quad E(\Phi(\cdot, t))\equiv E(\Phi(\cdot, 0)),\quad t \geq 0,$$
where the energy is defined as
   \begin{equation}\label{energy def0} E(\Phi(\cdot, t)) := \int_{\mathbb{R}^{d}} \left[ \frac{1}{2}|\nabla\Phi(\mathbf{x}, t)|^{2} + V(\mathbf{x})|\Phi(\mathbf{x}, t)|^{2} + \frac{\beta}{2}|\Phi(\mathbf{x}, t)|^{4} - \Omega \overline{\Phi} L_{z} \Phi (\mathbf{x}, t)\right] d\mathbf{x}.\end{equation}
A fundamentally important solution of (\ref{model}) is the ground state (GS), which is a stationary solution that minimizes the energy under a fixed mass \cite{aftalion2001vortices,BaoCai,bao2004computing}, i.e., 
\begin{equation}\label{GS def 0}
\psi_{gs}:=\mathop{\arg\min}\limits_{\Vert\psi\Vert_2=const}E(\psi).
\end{equation}
It corresponds to stable coherent states in many applications. In particular, with the presence of the rotating effect, i.e., $\Omega>0$, 
quantized vortices would form various delicate patterns in GS \cite{fetter2009rotating}, leading to phase transitions \cite{BaoCai,3Dexample}.  
For computing GS, various state-of-the-art numerical methods based on classical discretizations have been proposed, including imaginary-time evolution/gradient-flow type methods \cite{Duboscq,bao2005ground,Besse,Xie,danaila2010new,Zhangyong}, constrained optimization methods \cite{antoine2017efficient,Protas,TangZhang}, and eigenvalue methods \cite{altmann2021j,cances2010numerical,HenningSIRev}. For a study of the energy landscape, we refer to \cite{Lei}.  Although powerful and mathematically elegant, these traditional methods are primarily suited to low-dimensional scenarios, as their computational cost becomes prohibitively high in higher dimensions. Moreover, they are restricted to the computation of GS under fixed given physical setup. For any changes in the physical setup, e.g., a new value of $\Omega$ or a new potential $V(\mathbf{x})$, one will need to re-run the computer program for the corresponding GS. Such repeated computations are costly, particularly in the rotating case. For studies on the action GS, we refer to \cite{LiuZhao2,LiuZhao1}.


In recent years, deep neural networks (DNNs) have emerged as powerful tools for scientific computing \cite{Ritz,FNO,lu2021learning,raissi2019physics}. They offer feasibility for high-dimensional problem and  superior generalization capabilities with respect to physical conditions.
Along the related research for GS of BEC,  the works \cite{bai2025rapid,bai2021,Bakthavatchalam,Liang} have considered effective DNN approximations of GS and parameter generations based on supervised training, where the training data for labeling are obtained from traditional numerical methods such as gradient flow.
The requirement of simulation data can indeed be expensive. To enable unsupervised training, 
Bao et al. recently \cite{bao2025computing} developed a normalized-DNN (N-DNN) framework for computing GS by directly working on the energy functional with DNN normalized with respect to mass. 
So far, all the aforementioned existing AI approaches are only for non-rotating GPE, i.e., $\Omega=0$ in (\ref{model}) or (\ref{energy def0}), corresponding to one-dimensional problem and/or relatively  simple coherent structure in GS.  As a matter of fact, the study of GS in the rotating case is more challenging than in the non-rotating case, due to the presence of quantized vortices. It calls for higher accuracy to capture the delicate structure and robustness to the transition of phase that is sensitive to the value of $\Omega.$

In this work, we continue the work of \cite{bao2025computing} by developing an unsupervised learning framework to effectively compute GS (\ref{GS def 0}) of rotating BEC, covering generalization with respect to the rotating speed $\Omega$. 
Firstly, we will point out a limitation of N-DNN, which can affect the accuracy of GS. To address this issue, we will propose a new `normalized' loss as an alternative approach to maintain mass, while keeping the optimization unconstrained and the training unsupervised. A numerical comparison will confirm its improvement on accuracy. Secondly, to guide the training towards the right vortex  phase for GS, we introduce a novel training strategy named virtual rotation acceleration (VRA). Numerical experiments will show that without VRA, the usual training would fail, leading to incorrect GS predictions with the false number of vortices and pattern. In combination, the proposed strategies  form an accurate and efficient method for predicting GS. 
Finally, based on the proposed unsupervised learning method and a distillation technique, we will establish an operator network that maps from a physical parameter
like $\Omega$ to GS. It can offer a direct and fast prediction of GS with the right phase for all $\mathbf{x}$ and for all $\Omega$ in the prescribed domain that across vortex phases.  
As applications, the operator network is then used to an inverse problem that predicts the physical parameter based on an observed energy or solution image.

The rest of the paper is organized as follows. In Section \ref{sec:2}, we first give some preliminaries of GS and then present our key strategies for computation, followed by numerical experiments for validation. In Section \ref{generalization}, we give the operator network for generalization of physical parameters and present the applications. Some concluding remarks are drawn in Section \ref{conclusion}. All the numerical experiments in this paper are programmed in PyTorch sequentially and conducted on a desktop with an AMD Ryzen 7 9800X3D processor. The codes are available at \url{https://github.com/kongzhizhong/VRA}.

\section{Capturing GS with  vortices}\label{sec:2}
In this section, we present the methodology for employing a neural network to solve the GS of rotating GPE \eqref{model} under specified physical conditions.

\subsection{Preliminaries}
We give a brief review of some pre-knowledge for GS. The stationary solution of (\ref{model}) reads
\begin{equation}
    \Phi(\mathbf{x},t)=\psi(\mathbf{x})\mathrm{e}^{-i\mu t},\label{bian}
\end{equation}
where $\mu$ is interpreted as the chemical potential of the condensate. Substituting (\ref{bian}) into (\ref{model}) gives the following equation for the pair $(\mu,\Psi)$:
\begin{equation}
\mu\,\psi(\mathbf{x}) = -\frac{1}{2}\Delta\psi(\mathbf{x}) + V(\mathbf{x})\psi(\mathbf{x}) + \beta|\psi(\mathbf{x})|^{2}\psi(\mathbf{x})-\Omega L_z\psi(\mathbf{x}), \quad \mathbf{x} \in \mathbb{R}^{d},\label{eig}
\end{equation}
under the mass constraint, which is assumed to be normalized (w.l.o.g.)
\begin{equation}\label{eq:normalization}
    \|\psi\|_{2}^{2} := \int_{\mathbb{R}^{d}}|\psi(\mathbf{x})|^{2}\,d\mathbf{x} = 1.
\end{equation}
It is clear that \eqref{eig} is a nonlinear eigenvalue problem, and among all the solution pairs, the one of the most physical interest is the GS that minimizes the energy (\ref{energy def0}). 
%
More precisely, the GS solution denoted $\psi_{gs}$ is mathematically defined as the minimizer of the following constraint optimization problem:
\begin{equation}\label{GS def}
\psi_{gs}:=\mathop{\arg\min}\limits_{\Vert\psi\Vert_2=1}E(\psi).
\end{equation}

It is direct to see that $\psi_{gs}$ is an eigenfunction of \eqref{eig} with the Lagrangian multiplier $\mu=E(\psi_{gs}) + \int_{\mathbb{R}^{d}}\frac{\beta}{2}|\psi_{gs}|^{4}d\mathbf{x}$.  
The trapping potential $V(\mathbf{x})$ in this work will be fixed as the following harmonic oscillator which is the most concerned in applications, 
\begin{equation}
    V(\mathbf{x}) = 
    \begin{cases}
        \frac{1}{2}\left(\gamma_{x}^{2} x^{2} + \gamma_{y}^{2} y^{2}\right), & d = 2, \\
\frac{1}{2}\left(\gamma_{x}^{2} x^{2} + \gamma_{y}^{2} y^{2} + \gamma_{z}^{2} z^{2}\right), & d = 3,
    \end{cases}\label{potential}
\end{equation}
where $\gamma_x,\gamma_y,\gamma_z>0$ are the trap frequencies.
In this case, the existence of a GS for (\ref{GS def}) is theoretically guaranteed \cite{bao2005ground, seiringer2002gross} for  $\beta\geq0$ when $\vert\Omega\vert<\gamma_{xy}:=\min\{\gamma_x, \gamma_y\}$ in both 2D ($d=2$) and 3D ($d=3$) case. In this paper, we will obtain the GS for reference using the discrete normalized gradient flow \cite{BaoCai} or the preconditioned conjugate 
gradient method (PCG) \cite{PCG}.

\begin{figure}[h!]
    \centering
    \includegraphics[width=0.85\linewidth]{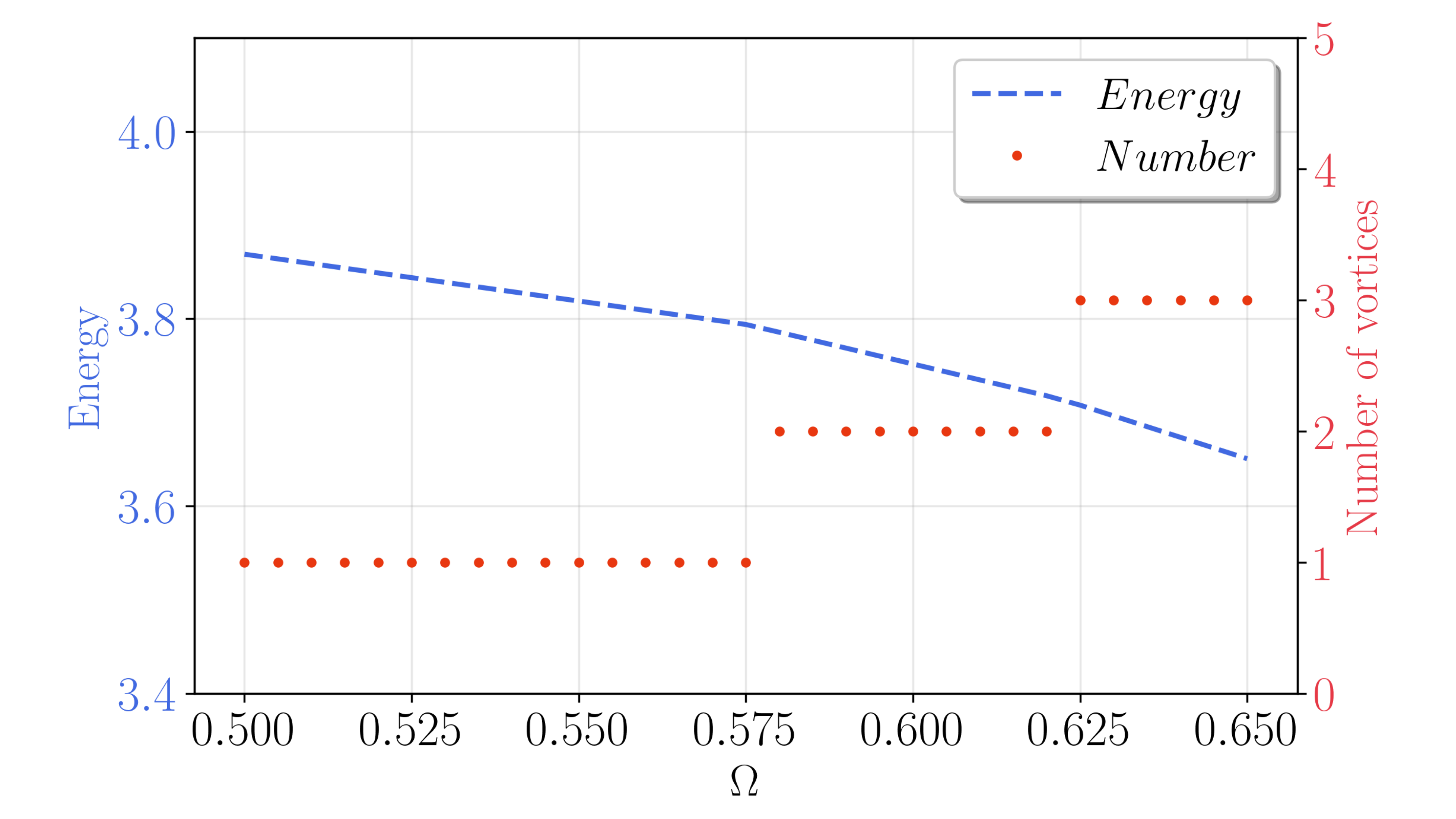}
    \caption{Total number of vortices and the energy of GS under different $\Omega$.}
    \label{quxian0}
\end{figure}

As $\Omega$ increases and exceeds a critical speed, quantized vortices that are localized phase singularities with
integer topological charge will gradually appear in GS. They can form attractive patterns and have been observed experimentally \cite{vortexexp}.  A notable fact is that as $\Omega$ increases, 
the energy of GS decreases with increasing total number of vortices in GS. See Fig.~\ref{quxian0} for an illustration. This will play a guide role in our algorithm design.

\subsection{DNN with normalized loss}
To compute GS, the first challenge lies in ensuring the mass constraint. 
For more accurate and structure-preserving results, we require exact global satisfaction rather than approximate enforcement through penalty terms. To this end, \cite{bao2025computing} introduces Normalized DNN (N-DNN): $\psi_\theta/\|\psi_\theta\|_2$ with $\psi_\theta$ a standard fully connected neural network \cite{Hinton}, where a batch $L^2$-normalization layer is incorporated into the network architecture (see \cite{ChangWenZhao} for other types of normalization) to keep the mass constraint inherently and thus turns  (\ref{GS def}) into an unconstrained optimization. This approach has proven effective in identifying GS without rotation, while it could be limited by insufficient precision to capture delicate vortice structures in GS under the rotating effect (see Section \ref{sec:compare_normalization}). In fact, the N-DNN model may exhibit non-negligible errors in computing derivatives via the automatic differentiation if not specially implemented. Let us explain this issue in the following. 

Taking the gradient operator $\nabla$ as an example, for any differentiable function $\psi$ with $\|\psi\|_{2}>0$, the following relation holds at the continuous level:
\begin{equation}\label{eq:grad_relation}
    \nabla\left(\frac{\psi}{\Vert \psi\Vert_2}\right) =\frac{\nabla\psi}{\Vert \psi\Vert_2}.
\end{equation}
In practice, the integral in $\Vert \psi\Vert_2$ is approximated using some quadrature rule, e.g., the uniform grid or Monte-Carlo sampling:
   $$ \Vert \psi\Vert_2 \approx \Vert \psi\Vert_{l^2} := \sqrt{\frac{\vert G\vert}{N}\sum_{i=1}^{N}\vert\psi(\mathbf{x}_i)\vert^2},$$
where $\vert G\vert$ denotes the measure of the computational domain $G\subset\mathbb{R}^d$ and $\{\mathbf{x}_j\}_{j=1}^N$ represents the sample points $N$ drawn from $G$. The gradient in N-DNN is then computed through the left-hand side of \eqref{eq:grad_relation} with the discretized $\Vert \psi\Vert_2$, and note that here the sample points $\{\mathbf{x}_j\}_{j=1}^N$ for discretization coincide with the batch samples. Under the common automatic differentiation in the machine learning package, this leads to the following approximation of the gradient operator at each sample $\mathbf{x}_k$ of the training/prediction batch, denoted as  $\nabla_{\mathbf{x}_k}$:
\begin{equation}\label{eq:app_NDNN}
   \nabla\left(\frac{\psi}{\Vert \psi\Vert_2}\right) ({\mathbf{x}_k})\approx \nabla_{\mathbf{x}_k}\left(\frac{\psi}{\Vert \psi\Vert_{l^2}}\right) = 
   \frac{\nabla_{\mathbf{x}_k}\psi}{\Vert \psi\Vert_{l^2}} - \frac{\psi\nabla_{\mathbf{x}_k}\Vert \psi\Vert_{l^2}}{\Vert \psi\Vert_{l^2}^2}.
\end{equation}
The second term on the right-hand side of \eqref{eq:app_NDNN} arises from the batch normalization embedded in N-DNN, which introduces additional errors. This kind of error becomes more apparent in rotating BEC than in the non-rotating case \cite{bao2025computing}, due to the presence of angular momentum operator $L_z = -i(x\partial_y - y\partial_x)$ that brings more derivatives.

To mitigate this issue, here we consider directly discretizing the right-hand side of \eqref{eq:grad_relation}. This leads to a new normalization strategy: instead of normalizing the network, we employ a `normalized' loss function to achieve mass conservation in \eqref{eq:normalization}. Specifically, let $\Phi = \frac{\psi}{\|\psi\|_2}$ for some function $\psi(\mathbf{x})$ with $\|\psi\|_{2}>0$, the energy (\ref{energy def0}) becomes the `normalized' energy functional: 
    $$E=\int_{\mathbb{R}^d} \left[\frac{1}{2}\frac{\vert\nabla\psi\vert^2}{\Vert\psi\Vert_2^2}+V\frac{\vert\psi\vert^2}{\Vert\psi\Vert_2^2}+\frac{\beta}{2}\frac{\vert\psi\vert^4}{\Vert\psi\Vert_2^4}-
\Omega \frac{\overline{\psi} L_{z} \psi}{\Vert\psi\Vert_2^2}\right]d\mathbf{x}.$$
By parameterizing $\psi$ with a DNN function $\psi_\theta$ and discretizing the integral above, we end up with the following new loss function:
\begin{equation}
    \mathcal{L}_{\Omega}(\theta):=\frac{\vert G\vert}{N}\sum_{j=1}^N\left[ \frac{1}{2}\frac{|\nabla\psi_{\theta}(\mathbf{x}_j)|^{2}}{\Vert{\psi_{\theta}}\Vert_{l^2}^2} + V(\mathbf{x}_j)\frac{|\psi_{\theta}(\mathbf{x}_j)|^{2}}{\Vert{\psi_{\theta}}\Vert_{l^2}^2} + \frac{\beta}{2}\frac{|\psi_{\theta}(\mathbf{x}_j)|^{4}}{\Vert{\psi_{\theta}}\Vert_{l^2}^4} - \Omega \frac{\overline{\psi_{\theta}} L_{z} \psi_{\theta}(\mathbf{x}_j)}{\Vert{\psi_{\theta}}\Vert_{l^2}^2} \right].\label{loss1}
\end{equation}

\begin{remark}
The additional error term in \eqref{eq:app_NDNN} can be further expressed as$$\frac{\psi\nabla_{\mathbf{x}_k}\Vert \psi\Vert_{l^2}}{\Vert \psi\Vert_{l^2}^2} = \frac{\vert G\vert}{N}\cdot\frac{\psi(\mathbf{x}_k)\nabla_{\mathbf{x}_k}\vert\psi(\mathbf{x}_k)\vert}{\Vert\psi\Vert_{l^2}^3}.$$
For any given $\mathbf{x}_k$, we can formally see that the right side of the above equation tends to zero when $N \to \infty$. This implies that when the batch size $N$ becomes sufficiently large, the computation error induced by auto-differentiation in the N-DNN method could be neglected, while this certainly brings unnecessary cost. 
\end{remark}


\subsection{Virtual rotation acceleration (VRA): a key training strategy}
\label{subsection:VRA}

The presence of rotation results in delicate vortex structures in GS (see Section \ref{neural network}). As $\Omega$ varies, GS undergoes sudden structural/topological discontinuities with an increasing or decreasing total number of vortices, while the corresponding GS energy inherently exhibit continuity (see Fig.~\ref{quxian0}). Consequently, direct training through \eqref{loss1} often gets stuck in wrong local minima, failing to capture the correct GS. This phenomenon will be demonstrated in Section \ref{detail} and it forms a  major obstacle.  

As shown above, the energy of GS decreases monotonically as the rotation speed $\Omega$ increases. 
To get close to the right GS, we find it vital to create manually a stronger rotation force, leading to a virtually enlarged rotation speed.
The accelerated rotation will drug an initial state of (\ref{loss1}) to a steeper trajectory heading closer and faster to the ground truth. 
To this end, we introduce a virtual rotation speed $\Omega^*\geq\Omega$, which defines a dual energy/loss:
\begin{equation}
    \mathcal{L}_{\Omega^*}=\frac{\vert G\vert}{N}\sum_{j=1}^N\left[ \frac{1}{2}\frac{|\nabla\psi_{\theta}(\mathbf{x}_j)|^{2}}{\Vert{\psi_{\theta}}\Vert_{l^2}^2} + V(\mathbf{x}_j)\frac{|\psi_{\theta}(\mathbf{x}_j)|^{2}}{\Vert{\psi_{\theta}}\Vert_{l^2}^2} + \frac{\beta}{2}\frac{|\psi_{\theta}(\mathbf{x}_j)|^{4}}{\Vert{\psi_{\theta}}\Vert_{l^2}^4} - \Omega^*\frac{\overline{\psi_{\theta}} L_{z} \psi_{\theta}(\mathbf{x}_j)}{\Vert{\psi_{\theta}}\Vert_{l^2}^2} \right].\label{loss2}
\end{equation}
The training for $\psi_\theta$ proceeds using \eqref{loss2} at an early stage and then switches back to the original \eqref{loss1} for refinement to the true GS. We refer to such a training strategy as \emph{virtual rotation acceleration (VRA)}, and it is outlined in Algorithm (\ref{power}).

\begin{algorithm}[!ht]
    \renewcommand{\algorithmicrequire}{\textbf{Input:}}
	\renewcommand{\algorithmicensure}{\textbf{Output:}}
	\caption{VRA training for DNN approximation of GS}
    \label{power}
    \begin{algorithmic}[1] 
        \REQUIRE Given the parameters $\Omega$, physical domain $G$. 
	    \ENSURE Full trained DNN $\frac{\psi_{\theta}}{\Vert\psi_{\theta}\Vert_2}\approx\phi_{gs}$. 
        \STATE Sample points for training: $\{\mathbf{x}_i\}_{i=1}^N\subset G$;
        \STATE Virtual stage: early train the DNN $\psi_{\theta}$ according \eqref{loss2};\label{11}
        \STATE Pullback stage: finetune the $\psi_{\theta}$ by continuing training according to \eqref{loss1}. 
    \end{algorithmic}
\end{algorithm}

 Subsequent experiments will demonstrate the efficacy of this approach. Specifically, an effective rotation speed that we practically identified and used is 
 \begin{equation}\label{eq:VRA_omega}
      \Omega^* = \min\{\frac{1+\Omega^2}{2},\gamma_{xy}\}.
 \end{equation}
Notice that $\Omega^*$ defined above satisfies the aforementioned condition.

\subsection{Numerical studies}\label{neural network}
We demonstrate here the effectiveness of our proposed method through a series of numerical experiments. The numerical solution obtained using DNN will be denoted by $\psi_{\theta}$, and the reference GS will be denoted by $\psi_{gs}$.

The whole space problem \eqref{GS def} is truncated to a bounded region $G\subset \mathbb{R}^d$, since GS under confinement (\ref{potential}) is rapidly decaying in the far field. The computational domain is set and fixed to $G = [-6, 6]^2$ for 2D problems and $G = [-6, 6]^3$ for 3D problems, discretized by uniform grids of $64 \times 64$ and $64 \times 64 \times 64$ points, respectively. These grid points serve as the sample points for the loss functions \eqref{loss1} and \eqref{loss2}. For consistency, all neural network models in the study utilize identical structures and hyperparameters. The DNN architecture here comprises five hidden layers, each containing 64 neurons, with the sine function as the activation function. To minimize the variance in the output of the neural network, the final output layer employs the hyperbolic tangent function (tanh) to constrain the output range within the interval $[-1, 1]$. The initial learning rate is set to $lr = 0.004$, which then decays by a factor of 0.95 every 100 epochs to refine convergence, and we utilize the DNN solution trained under \eqref{loss1} with $\Omega = 0$ as initialization. 
For some large value $\Omega$, $lr = 0.005$ ($\Omega\geq0.65$ or $\gamma_y\leq0.8$ in the following examples) and the DNN solution from $\Omega = 0.5$ ($\Omega\geq0.69$ in the following examples) are adopted as initialization to speed up the training process.

\subsubsection{Necessity of VRA}\label{detail} 
We first illustrate the necessity and effectiveness of VRA in training for the right GS. 

\begin{figure}[ht]
	 \centering 
	\begin{minipage}{0.3\linewidth}
		\vspace{1pt}
		\centerline{\includegraphics[width=\textwidth]{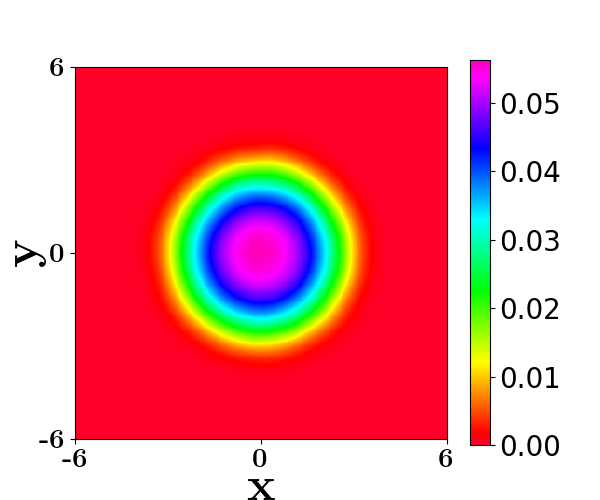}}
          
		\centerline{$epoch=0$}
	\end{minipage}
    \begin{minipage}{0.3\linewidth}
		\vspace{1pt}
		\centerline{\includegraphics[width=\textwidth]{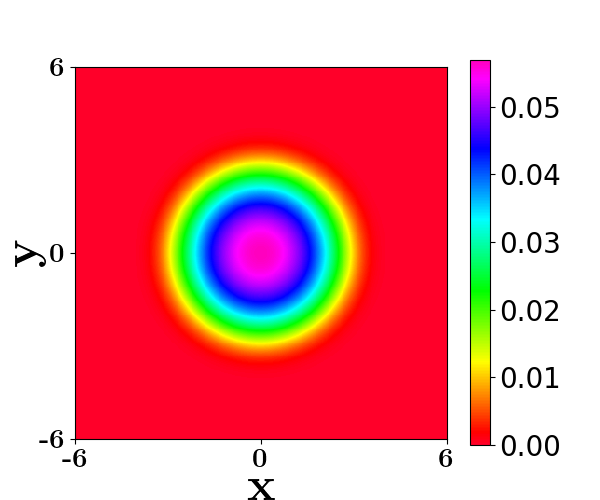}}
          
		\centerline{$epoch=2000$}
	\end{minipage}
	\begin{minipage}{0.3\linewidth}
		\vspace{1pt}
		\centerline{\includegraphics[width=\textwidth]{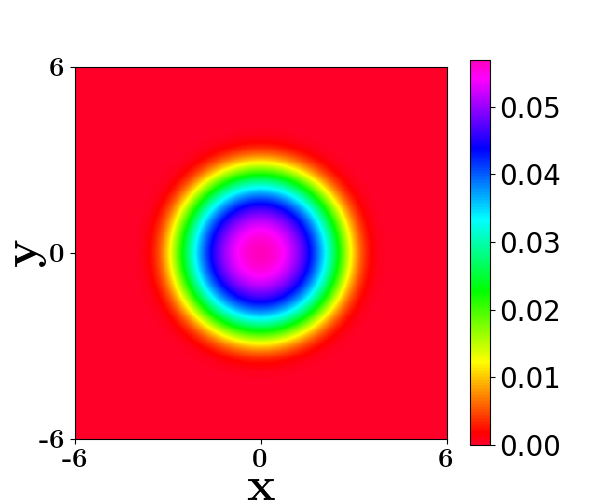}}
	 
		\centerline{$epoch=5000$}
	\end{minipage}

	\begin{minipage}{0.3\linewidth}
		\vspace{1pt}
		\centerline{\includegraphics[width=\textwidth]{fig/fine-tun/1.png}}
	 
		\centerline{$epoch=0$}
	\end{minipage}
    \begin{minipage}{0.3\linewidth}
		\vspace{1pt}
		\centerline{\includegraphics[width=\textwidth]{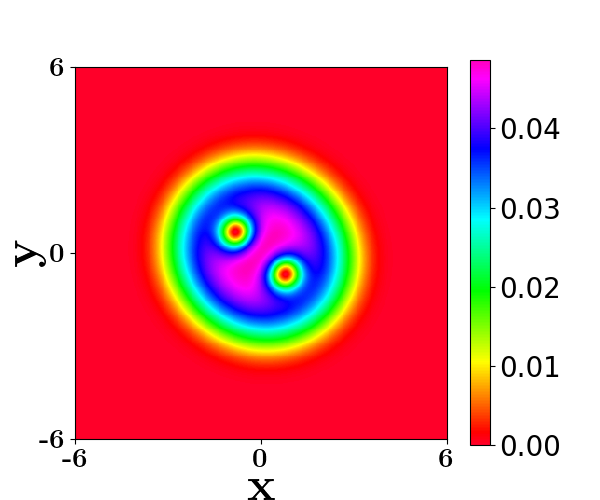}}
	 
		\centerline{$epoch=2000$}
	\end{minipage}
        \begin{minipage}{0.3\linewidth}
		\vspace{1pt}
		\centerline{\includegraphics[width=\textwidth]{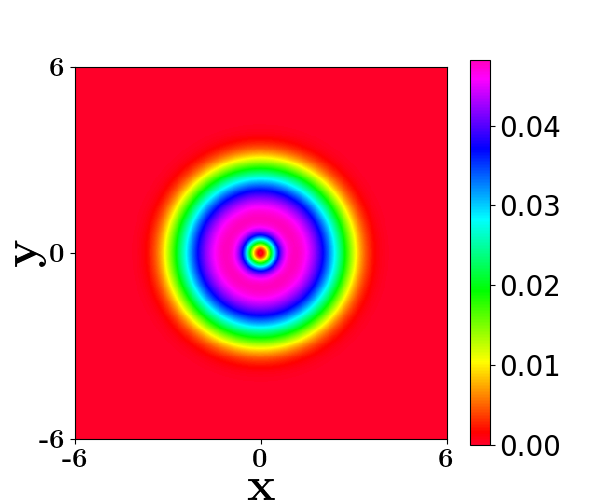}}
	 
		\centerline{$epoch=5000$}
	\end{minipage}

	\caption{Surface plot of $\vert\psi_{\theta}\vert^2$ under  $\Omega = 0.5$ and $\beta=100$ at different training epochs  without VRA  (1st row) or with VRA (2nd row).}
	\label{process}
\end{figure}

Training directly from the loss function $L_{\Omega}$ \eqref{loss1} often yields an inaccurate state. Specifically, the number of vortices obtained may be fewer than expected.
We take $\Omega=0.5$ and $\beta=100$ as an example 
to show this. In Fig.~\ref{process}, we plot the evolution of $|\psi_{\theta}|^2$ at different training epochs, where the first row depicts the results without VRA. The second row illustrates the training process with VRA, where $\psi_{\theta}$ is first trained using \eqref{loss2} for 3000 epochs, followed by fine-tuning with \eqref{loss1} for an additional 2000 epochs. 

As shown from the results in Fig.~\ref{process}, training without VRA converges to an incorrect state without vortices, and this cannot be fixed by tuning hyper-parameters or refining mesh. In contrast,  employing VRA enables the network to converge to the correct GS, capturing the right number of vortices. In particular, it is worth noting here the effect from VRA, where the enlarged virtual rotating speed first creates more (than the truth) vortices and then finetuning guides it back to the right GS. 

\begin{figure}[ht]
    \centering
    \includegraphics[width=0.9\linewidth]{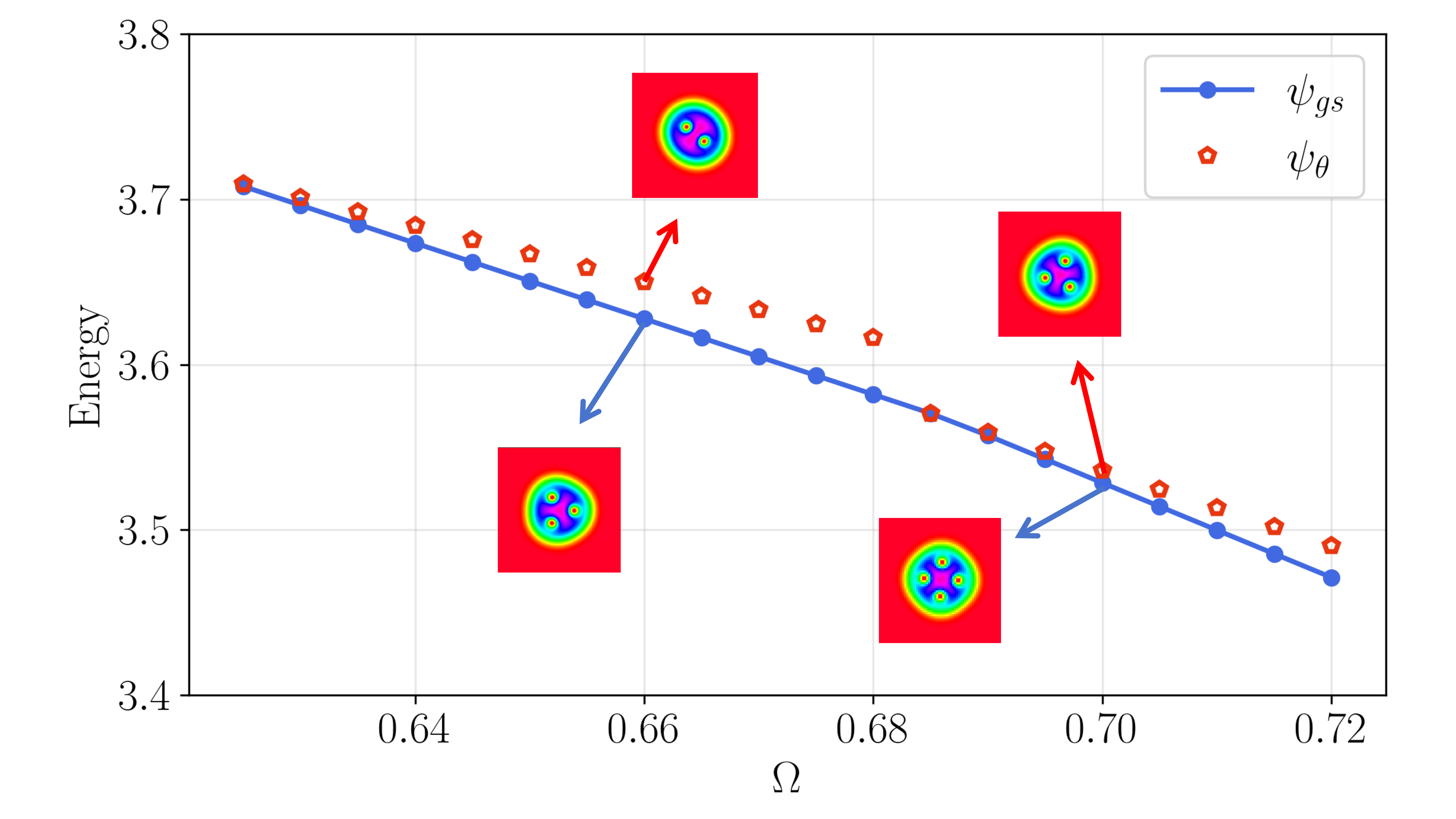}\\
\includegraphics[width=0.9\linewidth]{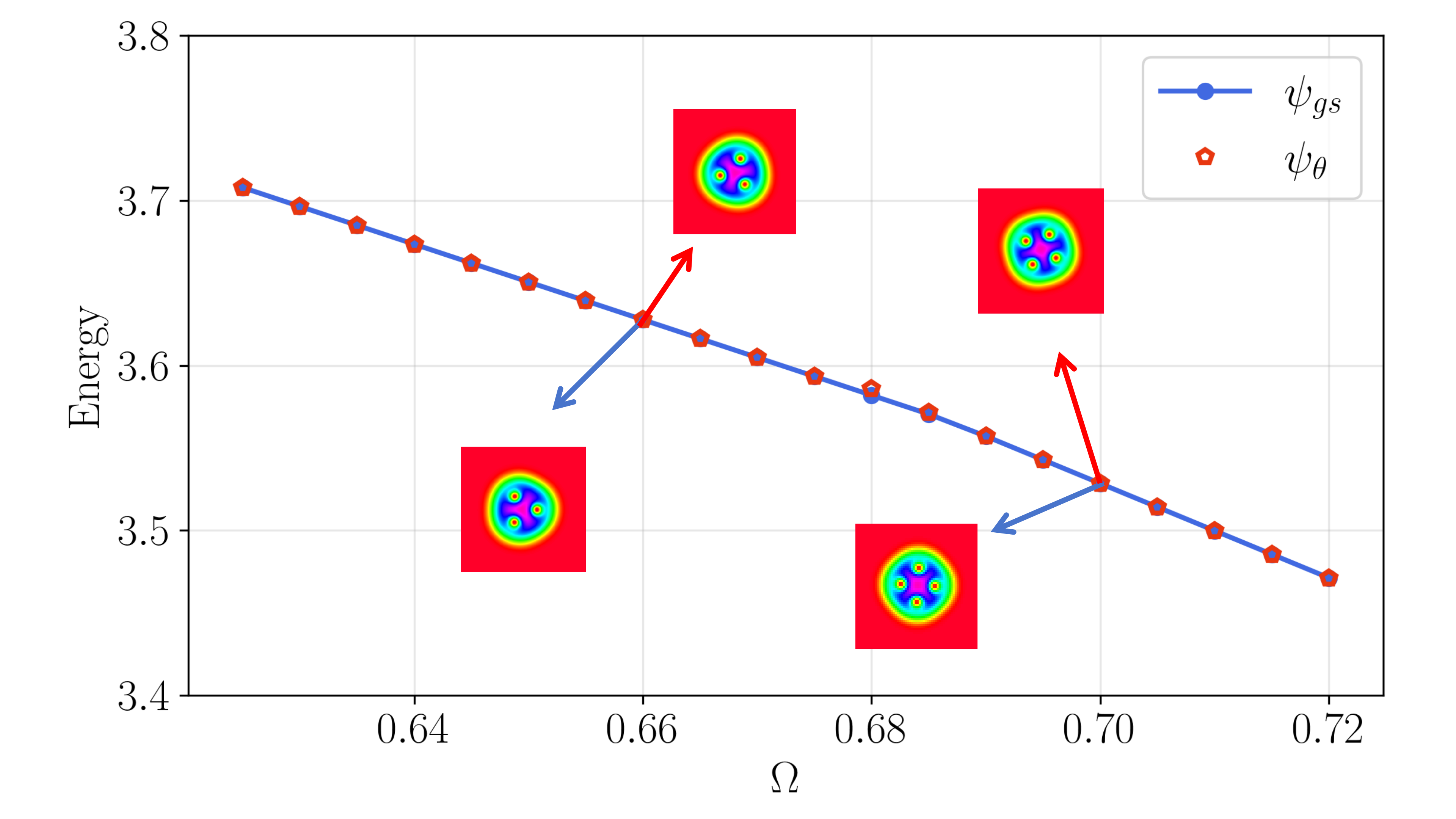}
    \caption{Energy of numerical solution obtained from training without VRA (1st row) and with VRA (2nd row) for different values of $\Omega$ under $\beta=100$.}
    \label{quxian}
\end{figure}

For further demonstration,  in the first row of Fig.~\ref{quxian}, we present the results of directly optimizing \eqref{loss1} without using VRA for several values of $\Omega$ that cross the phase transition of the vortex state. Compared to the true GS (blue points), direct optimization frequently yields higher converged energies (red points, over 5000 epochs) and a noticeable lagged phase transition in  $\psi_{\theta}$. For example, the  GSs of $\Omega = 0.66$ and $\Omega = 0.70$ which are highlighted in Fig.~\ref{quxian}, exhibit three and four vortices, respectively, whereas the corresponding numerical solutions produce only two and three vortices. The discrepancies in the energy and solution structure underscore the need for the proposed VRA strategy.

In the second row of Fig.~\ref{quxian}, we present the numerical results obtained using VRA in Algorithm \ref{power}: for each value of $\Omega$, we  pretrain $\psi_{\theta}$ with $\Omega^*$ given by \eqref{eq:VRA_omega} for 3000 steps, followed by fine-tuning with \eqref{loss1} for an additional 2000 steps. As shown, it yields the numerical solutions with structure and corresponding energy closely aligned with the true GS in most cases. Slight discrepancy occurs only near the critical values of $\Omega$ (e.g., $\Omega = 0.68$) for the phase transition of the vortex sate (e.g., from three vortices to four). 
This issue can be mitigated through further practical refinement, such as parameter-tuning and/or adjustment of the optimizer. 
Note that near such critical points of $\Omega$, traditional numerical methods also need special refinement for accuracy.  



\subsubsection{More validation of the method}\label{example_}
We perform more numerical experiments covering various scenarios here to further validate our method, i.e., Algorithm \ref{power}. The detailed setup of VRA here remains the same as before.  
We begin by exploring the case for a wider range of rotating speed. 

\begin{exmp}[GS in different rotating regimes]
Consider the 2D problem \eqref{GS def} with $\beta = 100$ and $V(\mathbf{x}) = \frac{1}{2}(x^2 + y^2)$. We consider the computation of GS for $\Omega=0.5,0.7,0.75,0.9$ (slow to fast rotation),  representing typical phases from single central vortex state to multiply complex vortices pattern. \label{exa1}
\end{exmp}

\begin{figure}[ht]
	 \centering 
	\begin{minipage}{0.48\linewidth}
		\vspace{1pt}
		\centerline{\includegraphics[width=\textwidth]{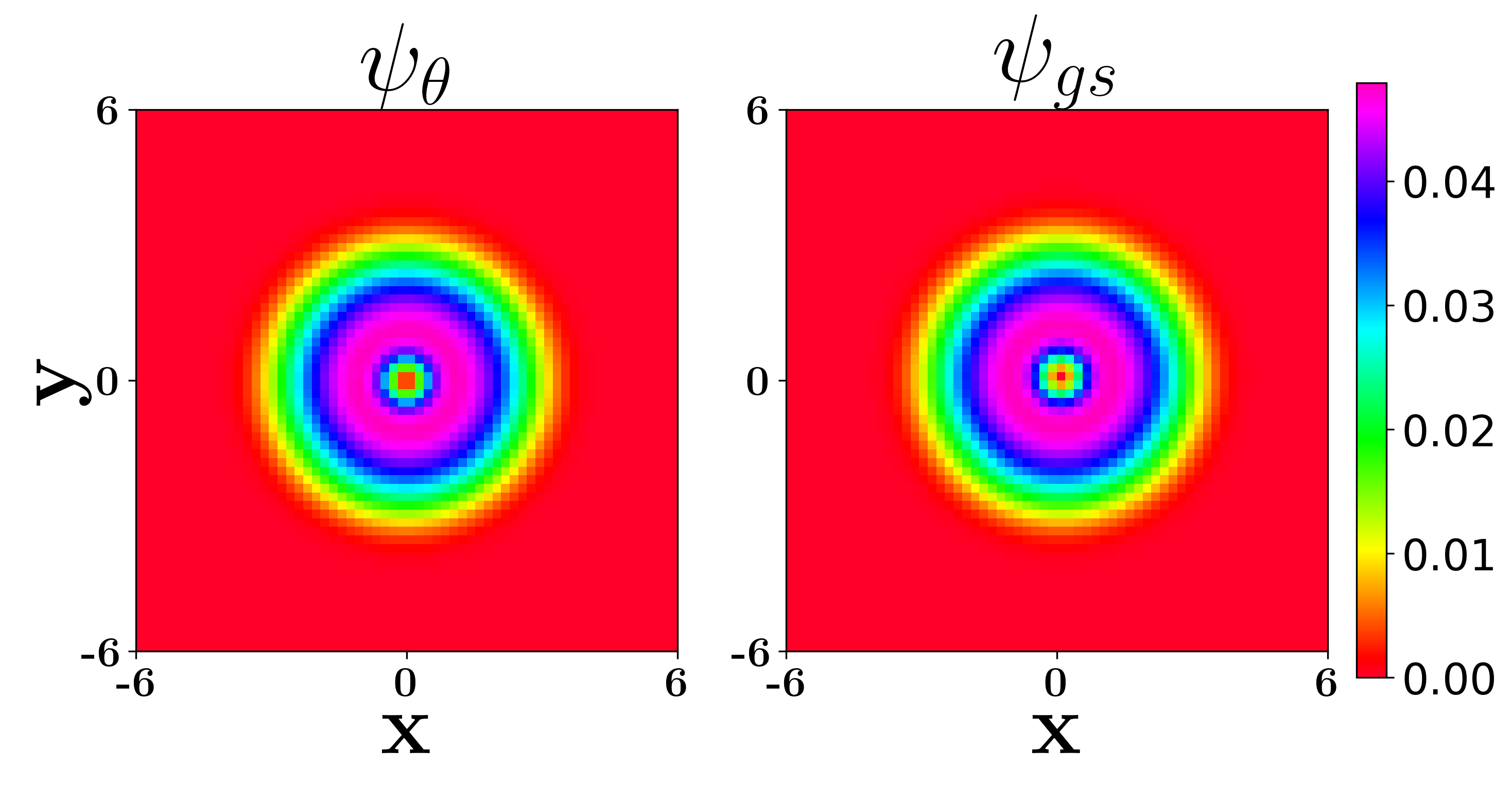}}
          
		\centerline{$\Omega=0.5$}
	\end{minipage}
	\begin{minipage}{0.48\linewidth}
		\vspace{1pt}
		\centerline{\includegraphics[width=\textwidth]{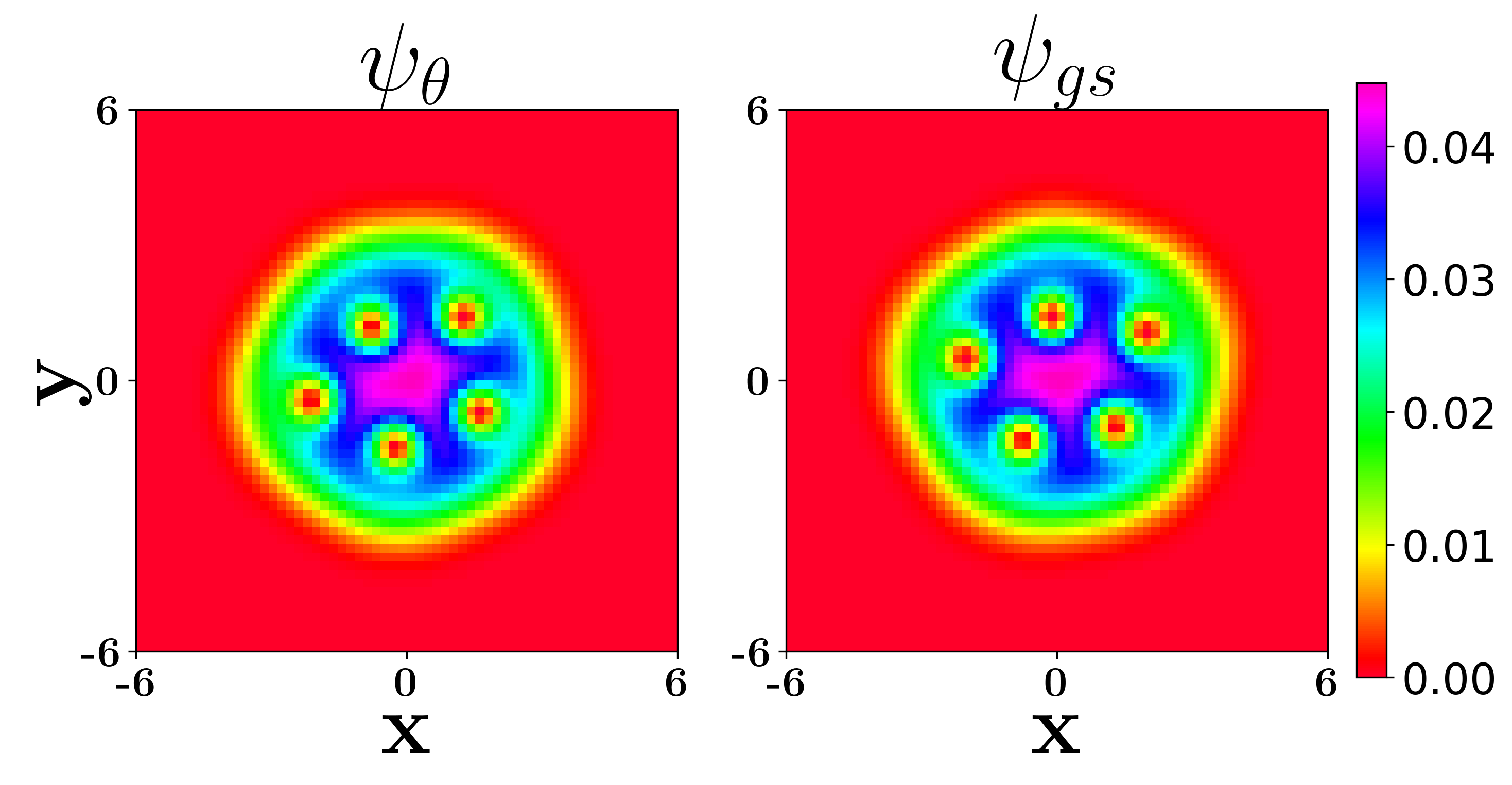}}
	 
		\centerline{$\Omega=0.75$}
	\end{minipage}

	\begin{minipage}{0.48\linewidth}
		\vspace{1pt}
		\centerline{\includegraphics[width=\textwidth]{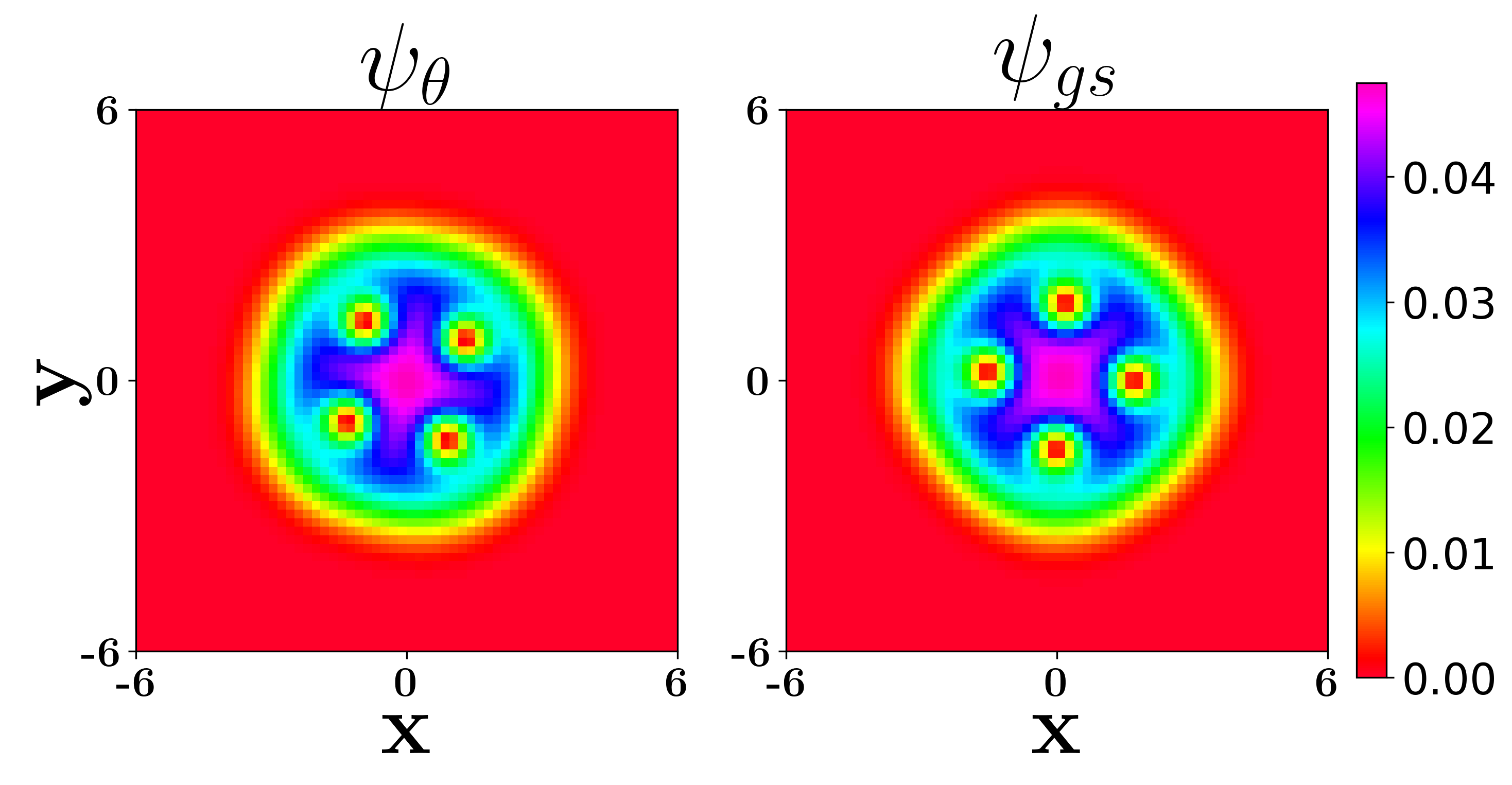}}
	 
		\centerline{$\Omega=0.7$}
	\end{minipage}
        \begin{minipage}{0.48\linewidth}
		\vspace{1pt}
		\centerline{\includegraphics[width=\textwidth]{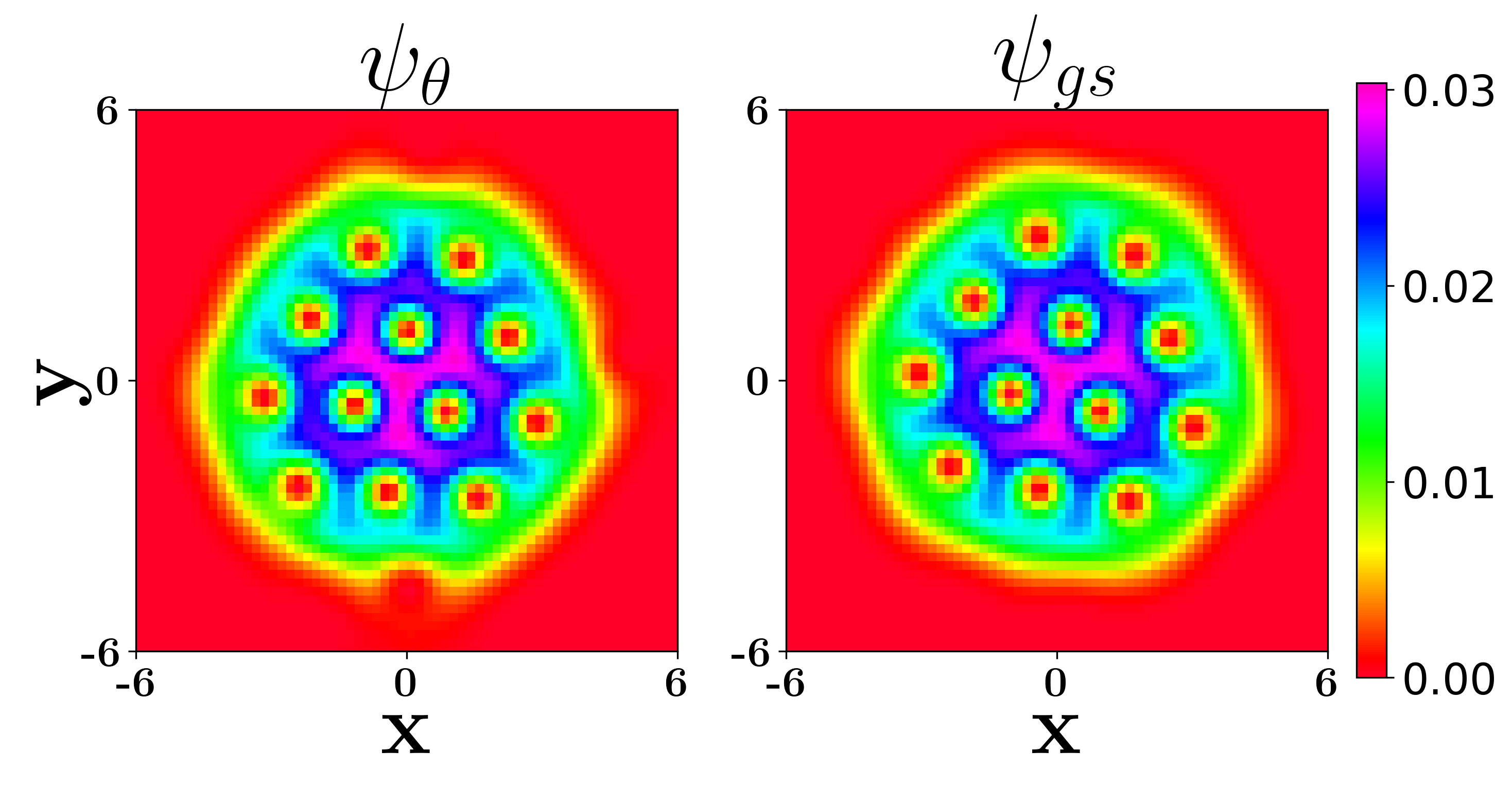}}
	 
		\centerline{$\Omega=0.9$}
	\end{minipage}
 
	\caption{Surface plots of  $\vert\psi_{\theta}\vert^2$ and $\vert\psi_{gs}\vert^2$  for different $\Omega$ in Example \ref{exa1}.}
	\label{fig1}
\end{figure}

We give the surface plot of  $\vert\psi_{\theta}\vert^2$ as well as the reference in
Fig.~\ref{fig1} for the different values of $\Omega$. Clearly, the results demonstrate that the numerical solutions obtained using the proposed method exhibit a high degree of consistency with GS in terms of vortex structure.

It should be noted that under a radial symmetric potential $V$, if $\psi$ is a GS, then any of its rotations in polar angle is still a GS with unchanged energy.  
This facts makes it difficult to give fair point-wise comparison metrics. Therefore, we assess the quality of our solution $\psi_{\theta}$, by comparing its  energy with that of the reference GS. For a consistent and accurate evaluation, we employ a unified spectral discretization on the grid to compute the energy functional for both solutions, denoted by $E_{\psi_{\theta}}$ and $E_{\psi_{gs}}$. 
The discrepancy is measured by the relative energy error:
\begin{equation}\label{energy err}
    Error = \frac{\vert E({\psi_{\theta}})-E({\psi_{gs}})\vert}{\vert{E({\psi_{gs}}})\vert},
\end{equation}
 and the results are quantified in Table \ref{table1}. It can be seen that high accuracy has been reached (about $10^{-5}$) from the slow to intermediate rotating regime. The accuracy only shows some notable reduction (but still reaches $10^{-3}$) in the fast rotating regime ($\Omega\geq0.9$) due to the more and more complex pattern in GS.


\begin{table}[ht!]
  \begin{center}
    \begin{tabular}{|c|c|c|c|c|c|c|c|} 
    \hline
      $\Omega$  & $0.6$ & $0.65$ & $0.7$ & $0.75$ & $0.8$ & $0.85$ & $0.9$\\
      \hline
      $\psi_{gs}$  & 3.7517  & 3.6506 & 3.5285 & 3.3820 & 3.1904 & 2.9557 & 2.6485\\
      \hline
      $\psi_{\theta}$ &  3.7518  & 3.6507 & 3.5286 & 3.3821 & 3.1905 & 2.9559 & 2.6530\\
      \hline
       Error  & $2.7E{-5}$ & $2.8E{-5}$ & $2.8E{-5}$ & $3.0E{-5}$ & $3.1E{-5}$& $6.8E{-5}$ & $1.7E{-3}$\\
      \hline
    \end{tabular} \label{table1}
    \caption{Energy values from $\psi_{\theta}$ and $\psi_{gs}$ and the error \eqref{energy err} for different $\Omega$ in Example \ref{exa1}.}
  \end{center}
\end{table}


We then investigate the scenario of a non-radial potential $V$ in  \eqref{potential}.

\begin{exmp}[GS under different confinement ratio]
    Take the 2D problem (\ref{GS def})  from Example \ref{exa1} but with $V(\mathbf{x})=\frac12(x^2+\gamma_y^2 y^2)$. 
    Now fix the rotational speed $\Omega = 0.5$ and we consider several values of trapping frequency $\gamma_y$. \label{exa2}
\end{exmp}

\begin{figure}[ht!]
	 \centering 
	\begin{minipage}{0.48\linewidth}
		\vspace{1pt}
		\centerline{\includegraphics[width=\textwidth]{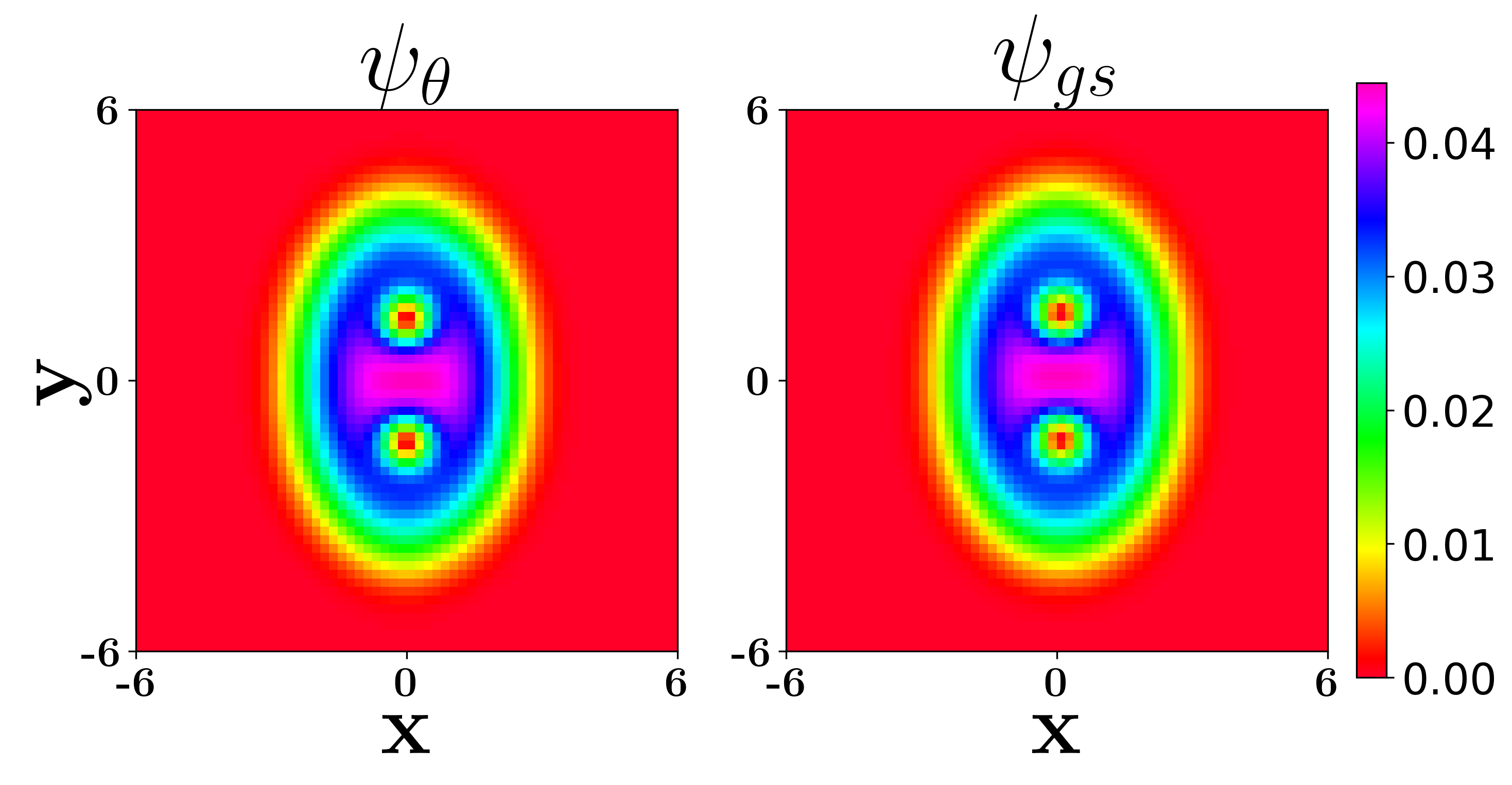}}
          
		\centerline{$\gamma_y=0.6$}
	\end{minipage}
	\begin{minipage}{0.48\linewidth}
		\vspace{1pt}
		\centerline{\includegraphics[width=\textwidth]{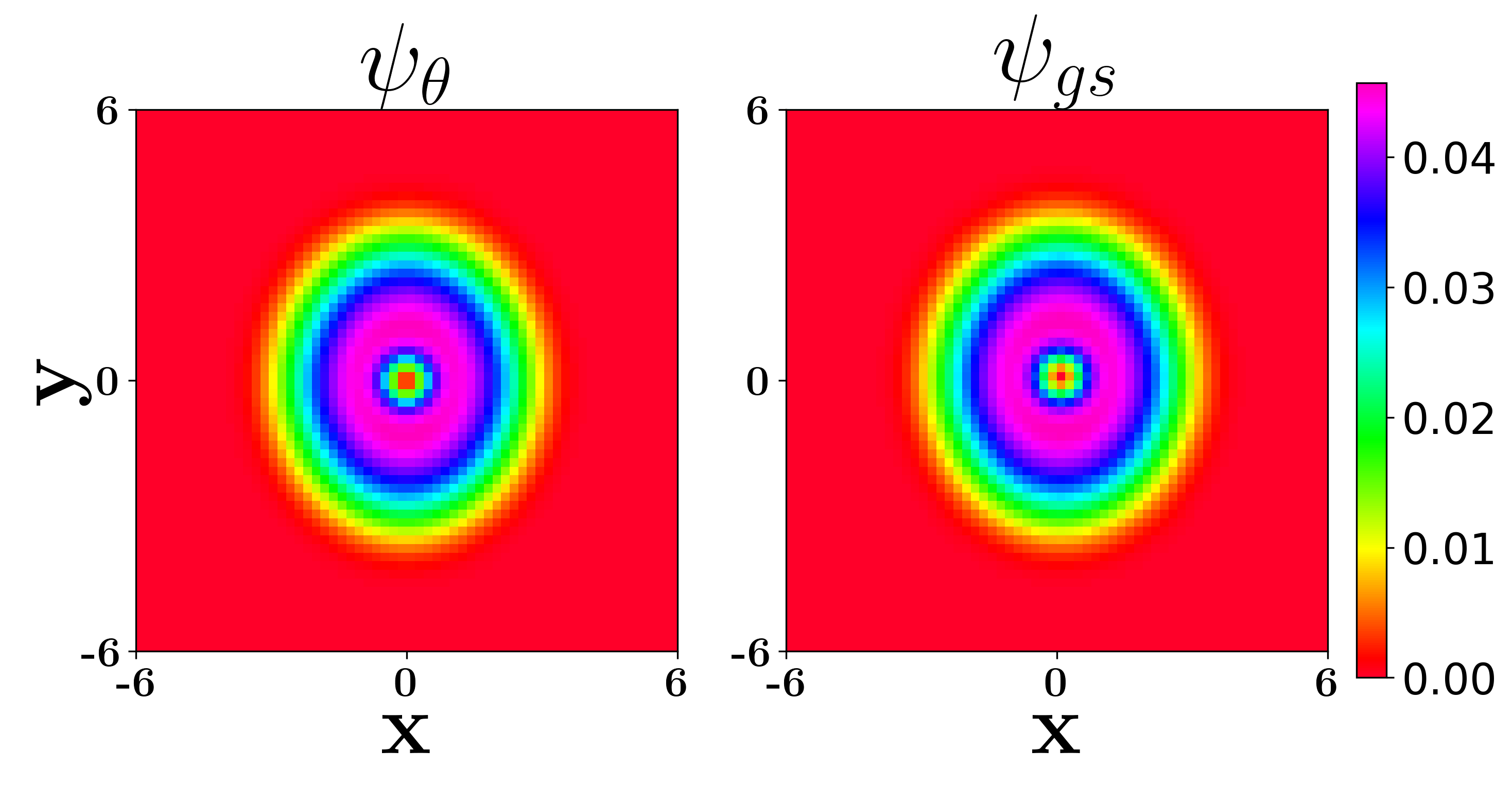}}
	 
		\centerline{$\gamma_y=0.8$}
	\end{minipage}

	\begin{minipage}{0.48\linewidth}
		\vspace{1pt}
		\centerline{\includegraphics[width=\textwidth]{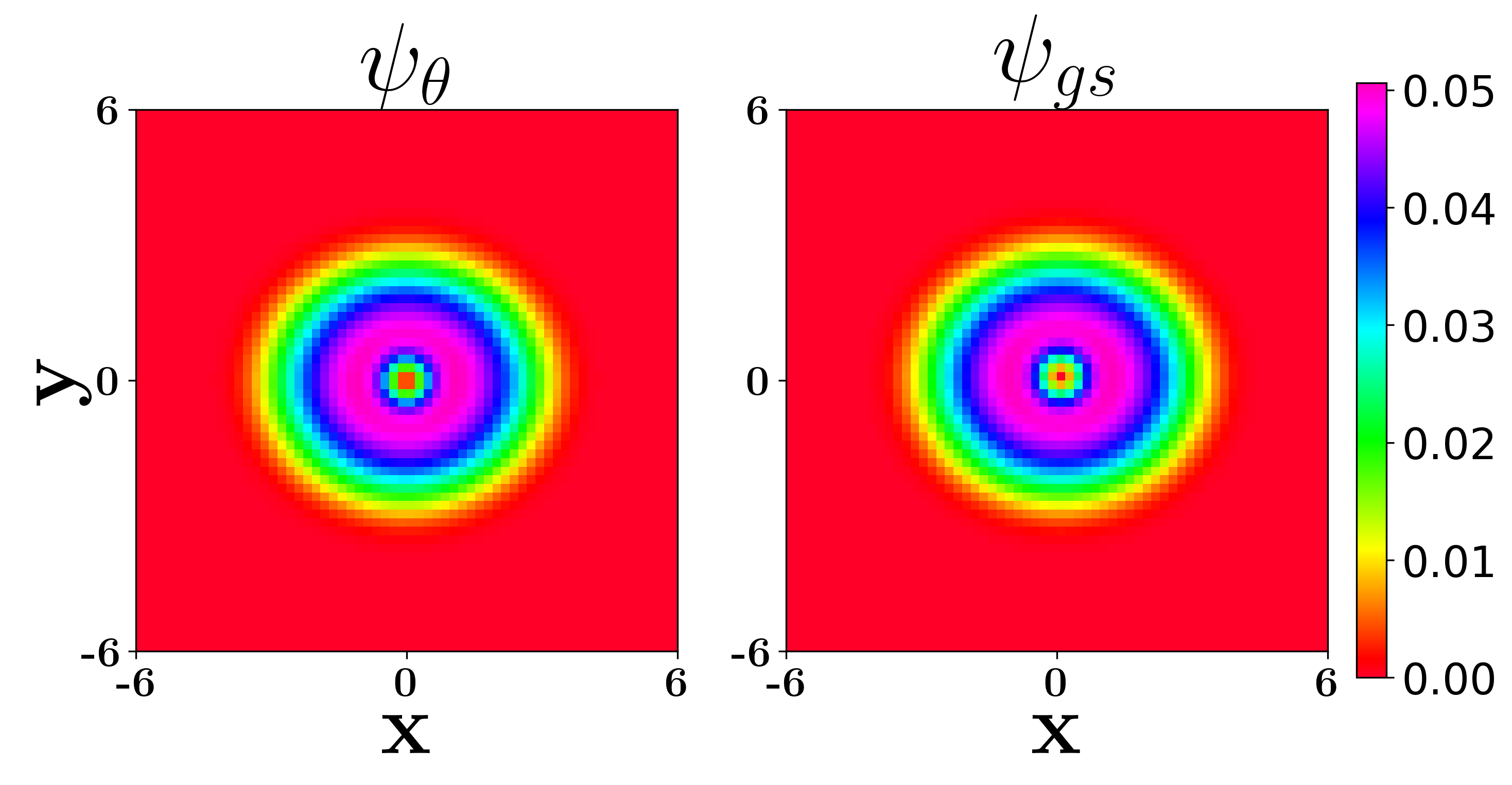}}
	 
		\centerline{$\gamma_y=1.2$}
	\end{minipage}
        \begin{minipage}{0.48\linewidth}
		\vspace{1pt}
		\centerline{\includegraphics[width=\textwidth]{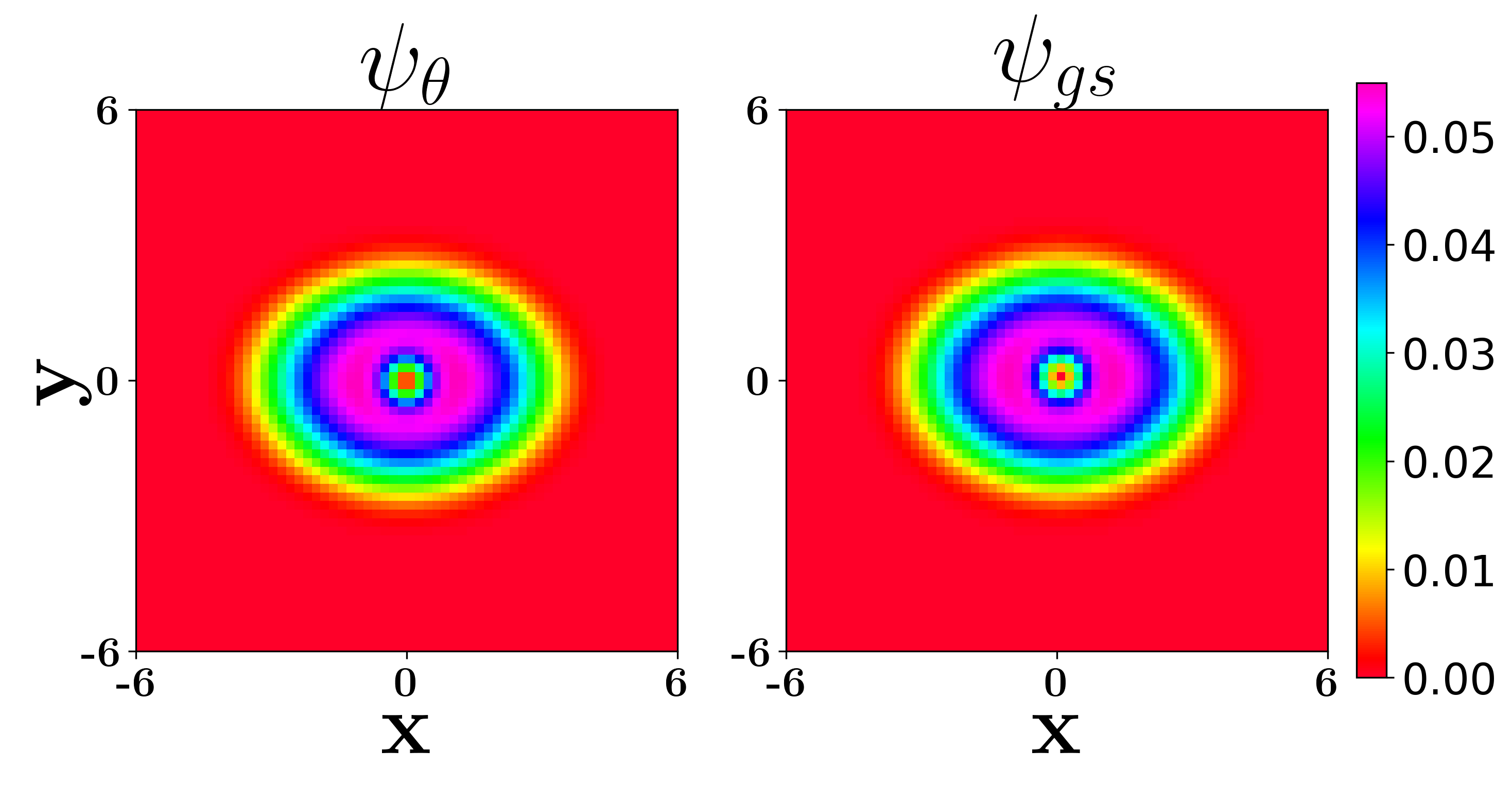}}
	 
		\centerline{$\gamma_y=1.6$}
	\end{minipage}

	\caption{Surface plots of $\vert\psi_{\theta}\vert^2$ and $\vert\psi_{gs}\vert^2$ for different $\gamma_y$ in Example \ref{exa2}.}
	\label{fig_v}
\end{figure}

The surface plots of the solutions from Example~\ref{exa2} are illustrated in Fig.~\ref{fig_v}. 
As can be seen, the changing value of $\gamma_y$ which reflects the anisotropy of confinement, can also cause different vortex phases in GS. The DNN solution correctly predicts the structure of the pattern. The corresponding quantitative metrics are presented in Table~\ref{table2}, where high precision is obtained.  The numerical outcomes here stay consistent with the findings as before.

\begin{table}[ht!]
  \begin{center}
    \begin{tabular}{|c|c|c|c|c|c|c|c|} 
    \hline
      $\gamma$ & $0.6$ & $0.8$ & $1.0$ & $1.2$ & $1.6$\\
      \hline
      $\psi_{gs}$ & 3.3051 & 3.6266  & 3.8689 & 4.0701 & 4.4000 \\
      \hline
      $\psi_{\theta}$ & 3.3056 & 3.6267  & 3.8690 & 4.0702 & 4.4001 \\
      \hline
      Error & $1.5E{-4}$ & $2.7E{-5}$  & $2.6E{-5}$  & $2.5E{-5}$ & $2.3E{-5}$ \\
      \hline
    \end{tabular}
    \caption{Energy values of $\psi_{\theta}$ and $\psi_{gs}$ and the error (\ref{energy err}) for different $\gamma$ in Example \ref{exa2}.}
    \label{table2}
  \end{center}
\end{table}

Together from the two examples, it is evident that our proposed DNN approach with normalized loss and VRA training strategy is capable of effectively and robustly capturing GS in different regimes of physical parameters of rotating BEC. Let us move on to a 3D test. 




\begin{exmp}[Three-dimensional case \cite{3Dexample}]\label{ex: 3D} 
    Consider the 3D GS problem (\ref{GS def}) with $\beta = 400$ and $V(\mathbf{x})=\frac12(x^2+y^2+z^2)$. We compute the solutions  for $\Omega = 0.6$ and $\Omega = 0.8$. 
\end{exmp}

\begin{figure}[ht]
     \centerline{$\Omega=0.6$:}
	\begin{minipage}{0.3\linewidth}
		\vspace{1pt}
		\centerline{\includegraphics[width=\textwidth]{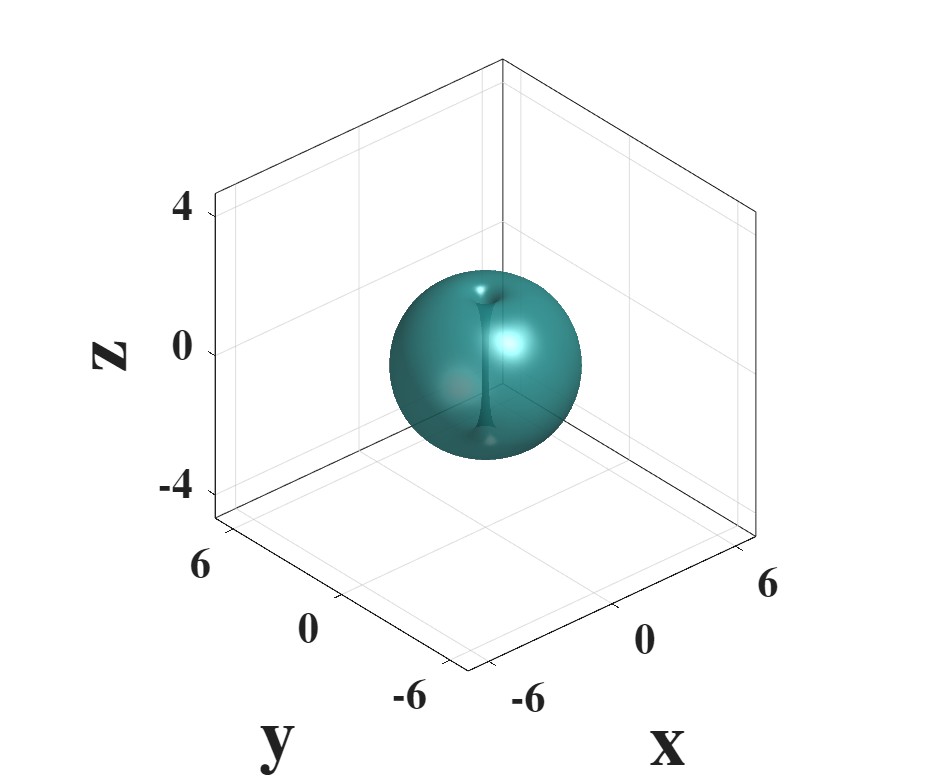}}

	\end{minipage}
    \begin{minipage}{0.3\linewidth}
		\vspace{1pt}
		\centerline{\includegraphics[width=\textwidth]{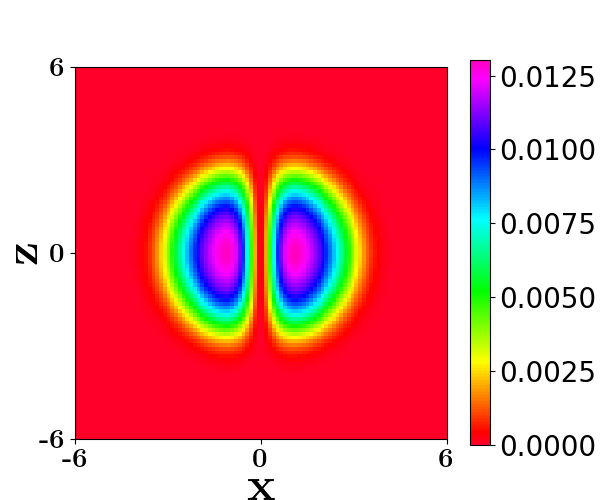}}

	\end{minipage}
	\begin{minipage}{0.3\linewidth}
		\vspace{1pt}
		\centerline{\includegraphics[width=\textwidth]{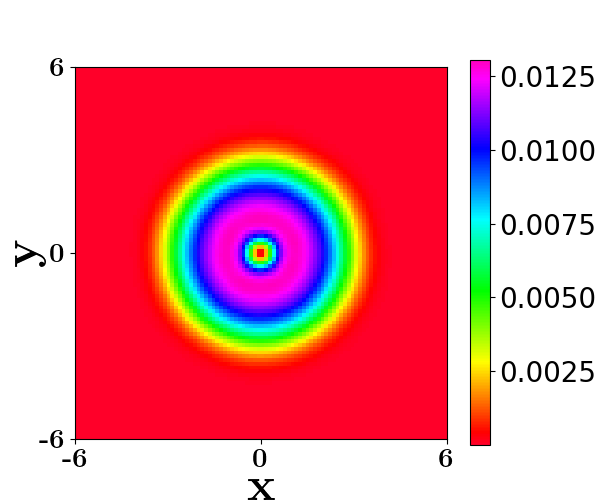}}

	\end{minipage}

\centerline{$\Omega=0.8$:}
    \begin{minipage}{0.3\linewidth}
		\vspace{1pt}
		\centerline{\includegraphics[width=\textwidth]{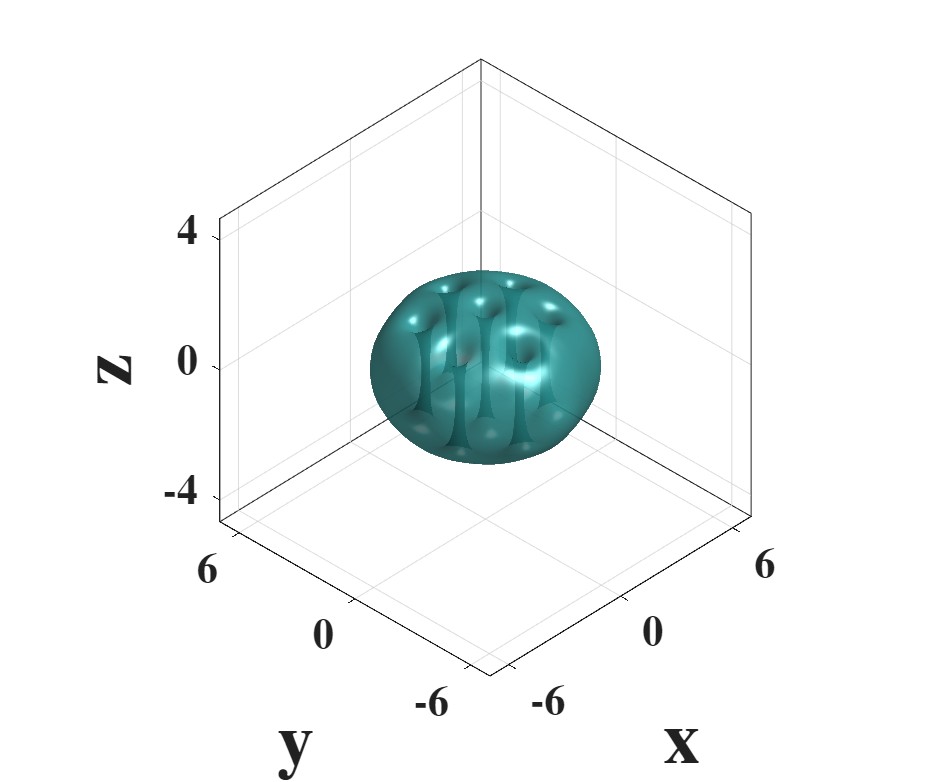}}

	\end{minipage}
    \begin{minipage}{0.3\linewidth}
		\vspace{1pt}
		\centerline{\includegraphics[width=\textwidth]{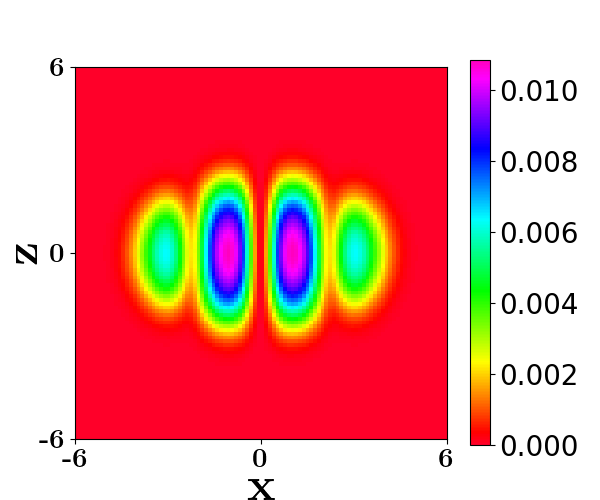}}

	\end{minipage}
	\begin{minipage}{0.3\linewidth}
		\vspace{1pt}
		\centerline{\includegraphics[width=\textwidth]{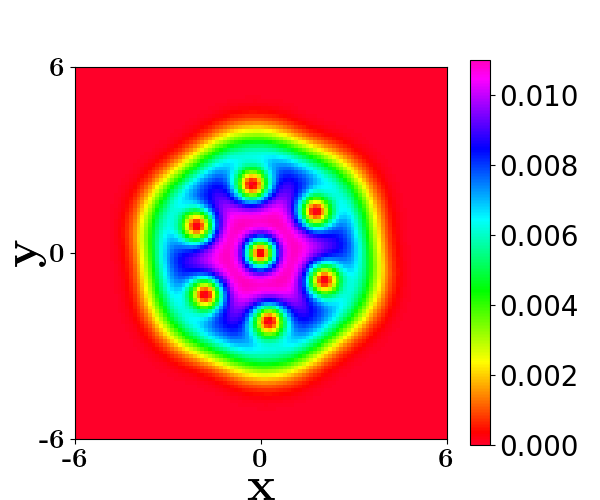}}

	\end{minipage}
 
	\caption{Numerical solution $|\psi_{\theta}|$ of Example \ref{ex: 3D} for $\Omega=0.6$  and $\Omega=0.8$: surface plot of $|\psi_{\theta}|^2$ (left column), the planar slice at $y = 0$ (middle column) and the planar slice at $z = 0$ (right column).}
	\label{3D_1}
\end{figure}

Here, for this example, we implement the VRA strategy for the initial 2500 epochs based on the loss function \eqref{loss2}, with the total training restricted to 4000 epochs. In Fig.~\ref{3D_1}, the computed solution $|\psi_\theta|$ from our approach is visualized using an iso-surface plot for function values exceeding a threshold of 0.001. As can be seen, for increasing rotating speed in the 3D case, the GS also undergoes a significant topological transition, leading to the formation of a complex, delicate, and multi-filamentary vortex state. It is worth highlighting here that 
\begin{itemize}
    \item For $\Omega=0.6$, the outcome of our method matches with that from \cite{3Dexample};
    \item For $\Omega=0.8$, the  solution $\psi_\theta$ of our approach reaches a steady state  different from that obtained by standard gradient flow in \cite{3Dexample}, where our solution in fact has a lower energy value $E(\psi_\theta)\approx3.8$. This numerical finding is then supported by the computational result of the PCG method \cite{antoine2017efficient}, where the same solution pattern with very close energy is produced, indicating that our $\psi_\theta$ is a more likely GS.
\end{itemize}

\begin{figure}[ht!]
    \centering
    \includegraphics[width=0.9\linewidth]{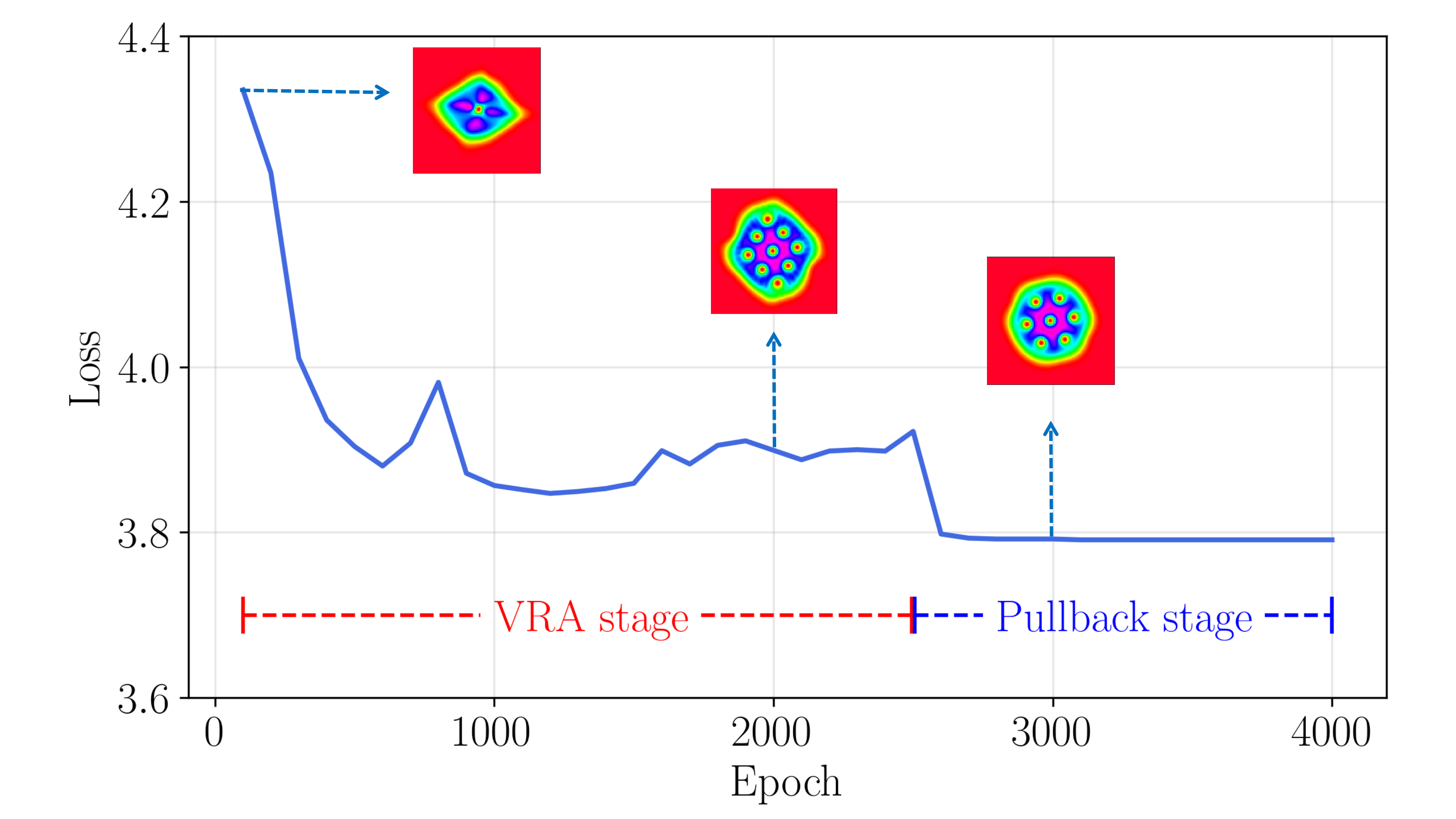}
    \caption{Training process of the 3D example: energy and evolution of 
     $\vert\psi_{\theta}\vert^2$ along iterations.}
    \label{3D:loss}
\end{figure}

The VRA training process is demonstrated in Fig.~\ref{3D:loss}, where the network solution is plotted at epochs 100, 2000, and 3000 within the 4000-epoch training run. 
The snapshots here reveal discernible differences between the two training stages, characterized by an overall trend where the number of vortices initially increases before decreasing, leading to the final convergence of the method.


The results obtained for this example all support that our proposed method remains effective in training and predicting GS in the 3D case. 

\subsubsection{Normalized loss vs normalized network}\label{sec:compare_normalization}
To highlight the practical improvement of the proposed normalized loss compared to the normalized network from the previous work, i.e., the N-DNN  \cite{bao2025computing}, here we conduct a comparison by using Example \ref{exa1}. 

\begin{figure}[ht]
	 \centering 
	\begin{minipage}{0.3\linewidth}
		\vspace{1pt}
		\centerline{\includegraphics[width=\textwidth]{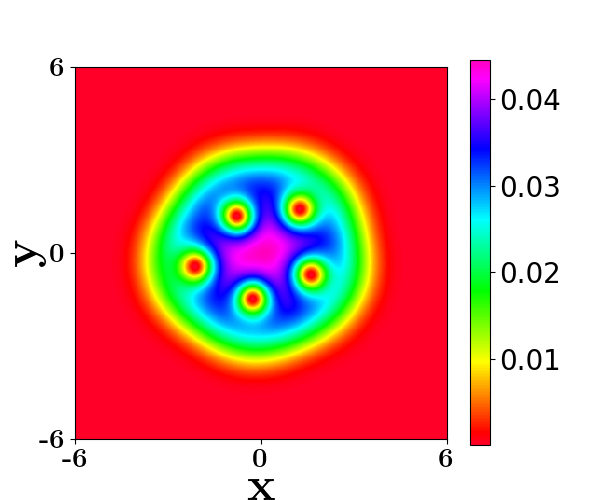}}
          
		\centerline{}
	\end{minipage}
    \begin{minipage}{0.3\linewidth}
		\vspace{1pt}
		\centerline{\includegraphics[width=\textwidth]{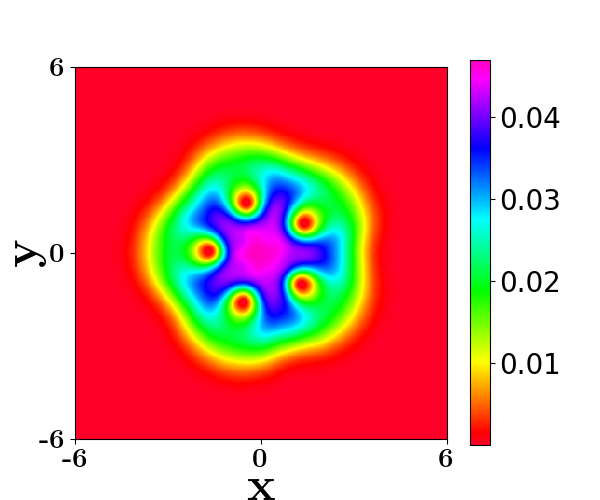}}
          
		\centerline{}
	\end{minipage}
    \caption{Surface plot of the numerical solution $\vert\psi_{\theta}\vert^2$ from the two different normalization techniques: normalized loss (left) and normalized network (right).}
	\label{norm}
\end{figure}

We take $\Omega=0.75$ and use the same computational setup, i.e., the same hyper-parameters, domain, sample points and VRA strategy, for both methods. The numerical solutions of the two approaches are shown in Fig.~\ref{norm}. The results confirm that the N-DNN method introduces a non-negligible error, leading to a higher final energy output (3.4108) compared to the current strategy (3.3821). The outcome of the normalized loss approach is consistent with the reference in Table \ref{table1}.  Although both solutions maintain the same number of vortices, their boundary configurations are markedly different, highlighting the critical influence of the normalization approach.

\subsubsection{Adaptive sampling}
As we have seen, the GS function of the rotating BEC is localized in space with high-gradient variations around the vortex core and potential well extrema. This makes traditional random or uniform sampling less efficient as they tend to allocate points equally. Here, to further improve the efficiency of our training, we consider an adaptive sampling strategy to optimize the allocation of computational resources by  assigning more sample points to regions with significant gradient changes and reducing points elsewhere. 

To achieve the goal, we utilize the $pygmsh$\footnote{An open-source package in Python available at https://pypi.org/project/pygmsh/.} function to generate the adaptive triangular mesh grid during training. 
It works by defining a size function for the mesh-size indication.
Here, we define
\begin{equation}\label{size fun}
Size(\mathbf{x}):=\max(\epsilon,1/\mathcal{G}(\mathbf{x})),
\end{equation}
 where $0\leq\mathcal{G}(\mathbf{x})\leq1$ is a gradient density function and $\epsilon$ represents the lower bound of the sampling density. A lower value of $Size$ indicates more densely scattered points. For the given sample points $\{\mathbf{x}_i\}_{j=1}^N$, $\mathcal{G}$ can be set by interpolating (e.g., cubic B-spline) from $(\mathbf{x}_j, \Vert\nabla\widetilde{\psi_{\theta}}(\mathbf{x}_j)\Vert_2)$ with the
normalized (to enhance stability) gradient
\begin{equation*}
    \nabla\widetilde{\psi_{\theta}}(\mathbf{x}_j)=\frac{\nabla\psi_{\theta}(\mathbf{x}_j)}{\max_{j=1,...,N}\{\Vert{\nabla\psi_{\theta}(\mathbf{x}_j)}\Vert_2\}}.
\end{equation*}

We incorporate the proposed method into the pullback stage of Algorithm \ref{power} and evaluate it on Example \ref{exa1}, setting $\Omega=0.9$. More precisely, the resampling step is performed at epoch 4000 with $\epsilon=0.1$, after which training is resumed. The process begins by deriving $\mathcal{G}$ from the initial 64×64 uniform grid. Then, (\ref{size fun}) is evaluated at 1,000 points randomly sampled within the interior of the domain  under uniform distribution. These points are utilized to generate the final mesh with pygmsh, with the total number of nodes controlled by adjusting the mesh size at the boundary.

\begin{table}[ht!]
  \begin{center}
    \begin{tabular}{|c|c|c|c|c|c|c|} 
    \hline
    Method & \multicolumn{5}{c|}{Adaptive} & Uniform \\
    \hline
      Number & $1728$ & $1917$ & $2126$ & $2507$ & $2937$ & $4096$\\
      \hline
      Energy & 2.6532 & 2.6526  & 2.6513 & 2.6511 & 2.6510 & 2.6530
      \\
      \hline
      Error & $1.8E{-3}$ & $1.5E{-3}$ & $1.1E{-3}$ & $9.8E{-4}$ & $9.4E{-4}$ & $1.7E{-3}$\\
      \hline
    \end{tabular}
    \label{table3}
    \caption{Energy and accuracy of the solution from adaptive sampling with different number of points.}
  \end{center}
\end{table}

The results presented in Table \ref{table3} demonstrate that the adaptive sampling algorithm achieves higher precision with fewer sample points compared to a uniform and fixed grid. In particular, to reach the same level of accuracy, our adaptive strategy requires only half the number of points, highlighting a significant improvement in training efficiency.

\section{Generalization in physical parameters}\label{generalization}
A drawback of traditional methods is their inefficiency when dealing with varying parameters: a complete re-run is required for each new set of physical parameters. 
In this section, we give an unsupervised operator learning for a family of parametric GS problems by applying the methodology established in Section \ref{sec:2}. Specifically, our objective is to learn a single operator $\Psi_\theta$ from the physical parameter such as rotating speed or trapping frequency (e.g., $\gamma_y$ in \eqref{potential}) to the corresponding GS function, i.e., 
\begin{equation}
    \Psi_\theta(\omega)(\mathbf{x})\approx\psi_{gs}^{\omega}(\mathbf{x}), \quad\forall\,\omega\in U,\label{operate}
\end{equation}
where $\psi_{gs}^{\omega}$ represents the GS corresponding to some physical parameter $\omega$ from a set $U$. For convenience of subsequent discussion, we denote the interval $U_i\subset\mathbb{R}_{+}\cup\{0\}$ with $i\in\mathbb{N}$ as the set of the concerned physical parameter $\omega$ ($\Omega$ or $\gamma_y$) such that for any $\omega\in U_i$, the GS of \eqref{GS def} exhibits $i$ vortices.

It is noted (Fig. \ref{quxian0}) that an increase in rotational speed $\Omega$ can induce a phase transition in GS, characterized by a change in the number of vortices. Within the intervals where the vortex number remains constant, the angular momentum exhibits a continuous and linear response to $\Omega$ until a new critical velocity is reached~\cite{vortexexp}. A similar effect can be observed for varying $\gamma_y$. Therefore, the set $U_i$ is a simply connected and non-intersected interval for both $\Omega$ and $\gamma_y$.



\subsection{Piecewise trained DNN}\label{fixed}
One major advantage of DNN is its approximation ability which is insensitive to increase of the dimensionality of the problem. It is therefore 
natural and convenient to consider the physical parameter(s) as an additional dimension of input to the neural network. In such a sense, we can still use the DNN function as the approximation of GS with the input variable $(\mathbf{x},\omega)$, i.e., $\psi_\theta(\mathbf{x},\omega)\approx\psi_{gs}^\omega(\mathbf{x})$. To extend the method proposed in Section \ref{sec:2}, particularly with the VRA strategy when $\omega=\Omega$, we construct the following loss functions: for a set of parameters from the target domain $\{\Omega_j\}\subset U$,
\begin{equation}
  \mbox{Virtual stage: }   \mathcal{L}_*(\theta):=\frac{1}{M}\sum_{j=1}^M\mathcal{L}_{\Omega_j^*}(\theta),\quad 
  \mbox{Pullback stage: } \mathcal{L}(\theta):=\frac{1}{M}\sum_{j=1}^M\mathcal{L}_{\Omega_j}(\theta),
  \label{loss3}
\end{equation}
with $\mathcal{L}_{\Omega}^*(\theta)$ as  in (\ref{loss2}) and  $\mathcal{L}_{\Omega}(\theta)$ as  in (\ref{loss1}). 

Our first goal is to verify the generalization capability of our model on the parameter $\Omega$ in one phase within an ideal interval, i.e., $U\subset U^i$. 
In experiments below, we construct a training set of $64 \times 64 \times M$ samples consisting of uniformly selected $N$ spatial points $\{x_i\}_{i=1}^{64 \times 64}$ and $M$ parameter points $\{\Omega_j\}_{j=1}^M$ from the interval $U$.


\begin{exmp}[Generalizing $\Omega$ in one phase]
    Consider the 2D GS setup in Example \ref{exa1} with $\Omega\in U=[0.58, 0.62]\subset U_2$. Set $\Omega_1=0.58, \Omega_M=0.62$. 
    \label{omega_omega}
\end{exmp}

\begin{table}[ht!]
  \begin{center}
    \begin{tabular}{|c|c|c|c|} 
    \hline
     Energy of $\psi_{gs}$ & \multicolumn{3}{c|}{3.7516} \\
      \hline
      Number & $M=2$ & $M=4$ & $M=6$\\
      \hline
      Energy of $\psi_{\theta}$ & 3.7521 & 3.7520 & 3.7520 \\
      \hline
      Error & $1.33E{-4}$ & $1.07E{-4}$ & $1.07E{-4}$ \\
      \hline
    \end{tabular}
    \caption{Energy value of $\psi_{\theta}$ and the error (\ref{energy err}) for $\Omega=0.6$ and different $M$ in Example \ref{omega_omega}.}
    \label{table:generalize_w}
  \end{center}
\end{table}

\begin{figure}[ht!]
	 \centering 
	\begin{minipage}{0.3\linewidth}
		\vspace{1pt}
		\centerline{\includegraphics[width=\textwidth]{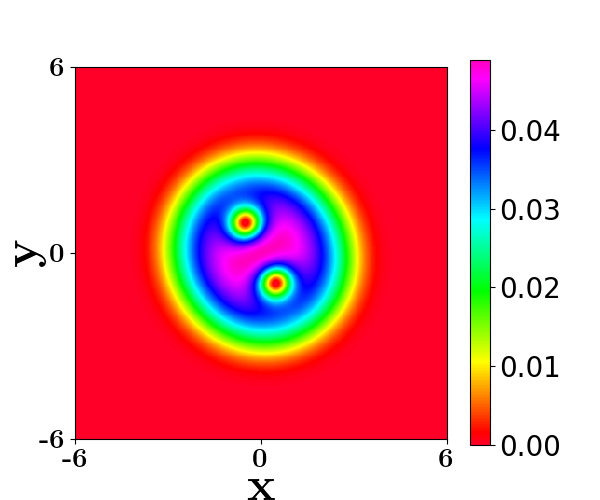}}
          
		\centerline{$\Omega=0.58$}
	\end{minipage}
	\begin{minipage}{0.3\linewidth}
		\vspace{1pt}
		\centerline{\includegraphics[width=\textwidth]{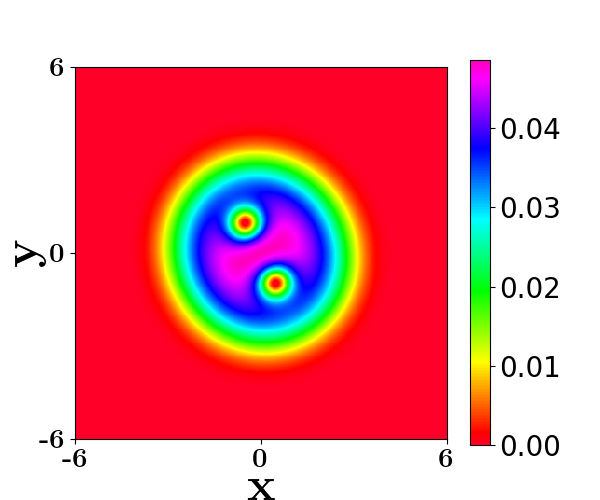}}
	 
		\centerline{$\Omega=0.6$}
	\end{minipage}
        \begin{minipage}{0.3\linewidth}
		\vspace{1pt}
		\centerline{\includegraphics[width=\textwidth]{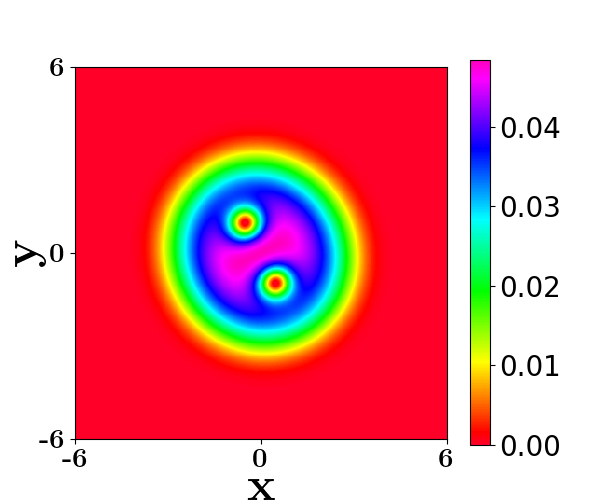}}
	 
		\centerline{$\Omega=0.62$}
	\end{minipage}

	\caption{Surface plot of  $|\psi_\theta|^2$ for different $\Omega\in U$ in Example \ref{omega_omega} with $M=2$.}
	\label{fig_o_suanzi}
\end{figure}

With several values of $M$, the accuracy is shown in Table \ref{table:generalize_w} for $\Omega=0.6$ (outside training set). The results validate our method and illustrate that even with the sparse sampling density in $\Omega$ in one phase, we get a high approximation precision. 
Fig.~\ref{fig_o_suanzi} depicts the numerical GS under the use of $M=2$,  where the vortex number remains constant throughout the parameter space. 

By changing the summation in \eqref{loss3} to that of the sampled $\{\gamma_{y,j}\}\subset U$ with $\Omega,\Omega^*$ fixed, similarly, we test and verify the generalization capability on the trapping frequency in the potential function which introduces anisotropy. 


\begin{exmp}[Generalizing $\gamma_y$ in one phase]
    Consider the 2D GS problem in Example \ref{exa2} with $\Omega=0.5$ and $\gamma_y\in U=[1, 1.6]\subset U_1$. Set $\gamma_{y,1}=1, \gamma_{y,M}=1.6$.\label{gamma_gamma} 
\end{exmp}

\begin{table}[ht!]
  \begin{center}
    \begin{tabular}{|c|c|c|c|} 
    \hline
    Energy of $\psi_{gs}$ & \multicolumn{3}{c|}{4.1601} \\
      \hline
       Number & $M=2$ & $M=4$ & $M=6$\\
      \hline
     Energy of $\psi_{\theta}$ & 4.1626 & 4.1606 & 4.1606 \\
      \hline
      Error & $4.81E{-4}$ & $1.20E{-4}$ & $1.20E{-4}$ \\
      \hline
    \end{tabular}
    \caption{Energy values of $\psi_{\theta}$ and the error (\ref{energy err}) for $\gamma_y=1.3$ and different $M$ in Example \ref{gamma_gamma}.}
    \label{table4}
  \end{center}
\end{table}

\begin{figure}[ht!]
	 \centering 
	\begin{minipage}{0.3\linewidth}
		\vspace{1pt}
		\centerline{\includegraphics[width=\textwidth]{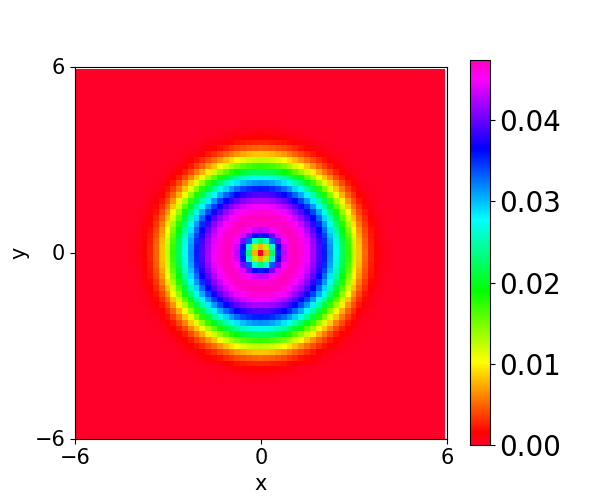}}
          
		\centerline{$\gamma_y=1$}
	\end{minipage}
	\begin{minipage}{0.3\linewidth}
		\vspace{1pt}
		\centerline{\includegraphics[width=\textwidth]{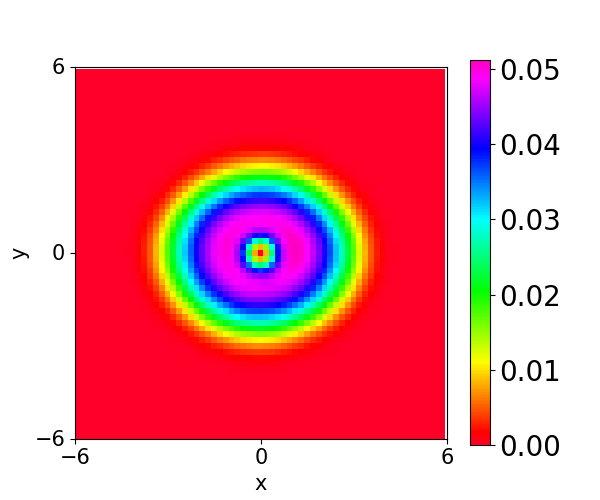}}
	 
		\centerline{$\gamma_y=1.3$}
	\end{minipage}
        \begin{minipage}{0.3\linewidth}
		\vspace{1pt}
		\centerline{\includegraphics[width=\textwidth]{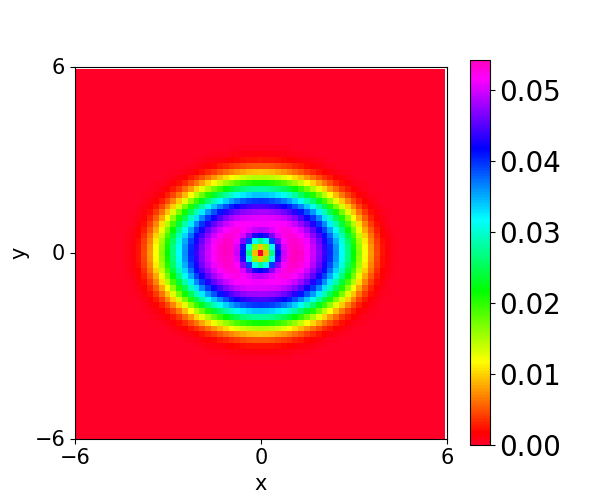}}
	 
		\centerline{$\gamma_y=1.6$}
	\end{minipage}

	\caption{Surface plot of $|\psi_\theta|^2$ for different $\gamma_y\in U$ in Example \ref{gamma_gamma} with $M=2$.}
	\label{fig_g_suanzi}
\end{figure}

Table \ref{table4} depicts the accuracy for $\gamma_y=1.3$ (outside training set) under different $M$, and 
Fig.~\ref{fig_g_suanzi} plots the numerical GS under $M=2$ where no phase transitions occur. Clearly,  the results support accurate generation also for the trapping 
frequency within a single phase. 

\begin{remark}
    For more general scenarios, such as directly predicting the GS under different forms of the potential $V(x)$ (beyond a single frequency parameter $\gamma_y$), advanced architectures such as FNO~\cite{FNO} can be employed in our training framework. This aspect will be explored in our future work.
\end{remark}

On the other hand, a direct generalization of $\omega$ across different phases will fail, as demonstrated by the following example.

\begin{exmp}[Failure of generalization across phases]\label{ex fail}
 Consider the generalization of $\Omega$ as an example. We train a  model via (\ref{loss3}) for $\Omega\in U=[0.5,0.6]$, and so $\Omega_1=0.5,\ \Omega_M=0.6$. The GS solution exhibits a single vortex at $\Omega=0.5$ and two vortices at $\Omega=0.6$.
\end{exmp}

\begin{figure}[ht]
	 \centering 
	\begin{minipage}{0.23\linewidth}
		\vspace{1pt}
		\centerline{\includegraphics[width=\textwidth]{fig/fine-tun/6.png}}
          
		\centerline{$\psi_{gs}$}
	\end{minipage}
    \begin{minipage}{0.23\linewidth}
		\vspace{1pt}
		\centerline{\includegraphics[width=\textwidth]{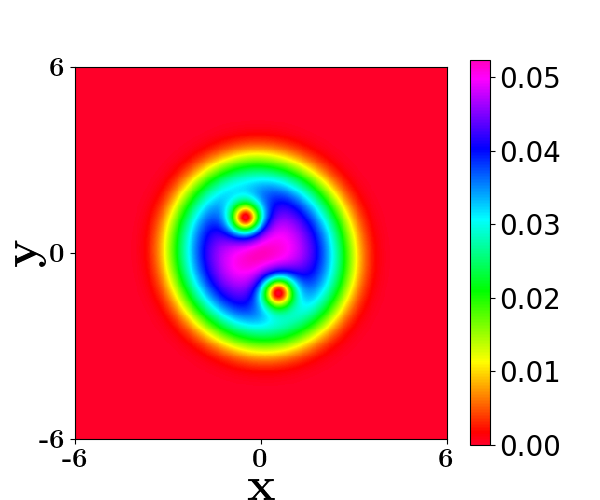}}
          \centerline{$\psi_\theta,\,M=2$}
	\end{minipage}
	\begin{minipage}{0.23\linewidth}
		\vspace{1pt}
\centerline{\includegraphics[width=\textwidth]{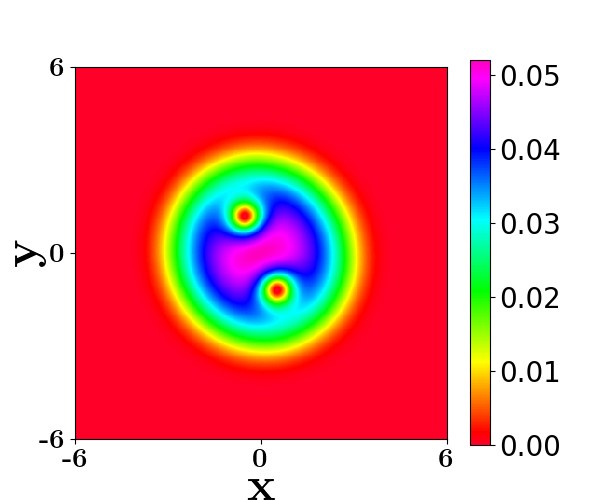}}
	 \centerline{$\psi_\theta,\,M=4$}
	\end{minipage}
        \begin{minipage}{0.23\linewidth}
		\vspace{1pt}	\centerline{\includegraphics[width=\textwidth]{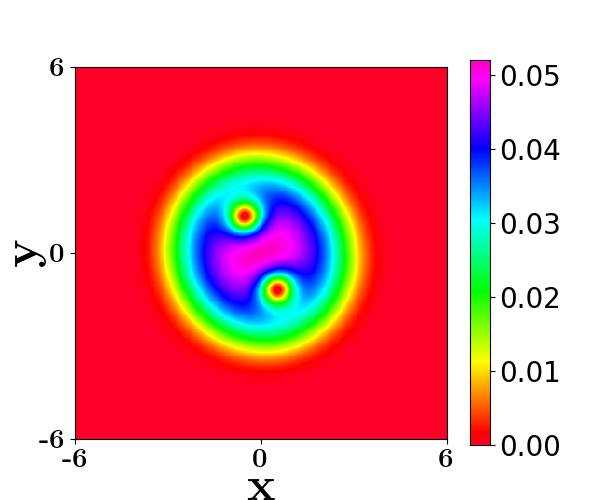}}
	 \centerline{$\psi_\theta,\,M=6$}
	\end{minipage}
	\caption{Surface plot of $\vert\psi_{\theta}\vert^2$ in Example \ref{ex fail} under $\Omega=0.5$ and different $M$.}
	\label{gene_22}
\end{figure}

As illustrated in Fig.~\ref{gene_22}, the directly across-phase 
trained model (underwent sufficient training epochs until loss convergence was achieved) failed to predict the left phase in GS, even at $\Omega=0.5$ from the training set. This is attributed to the critical phase-transition point $\Omega$ within the considered interval $U$, and the inherent continuity of neural network encounters difficulty to approximate sharp changes like discontinuity or high-frequency oscillations~\cite{sarma2024interface,selmic2002neural}. 

Consequently, for generalization of a parameter $\omega$ in the whole physically interested domain $U$, 
we are motivated to consider a piecewise trained DNN model for GS  within each stable regime where  phase transitions are not anticipated. More precisely, if  $U=\bigcup_j U_j$, we can train DNN operators $\{\psi^{(j)}\}_{j}$ separately on each $U_j$ as before. Note that pre-knowledge about the critical points for phase transition is available in the literature \cite{aftalion2001vortices,seiringer2002gross} or they can be determined in advance by accurate computations from Section \ref{sec:2}.  
The obtained $\{\psi^{(j)}\}_{j}$ together form a piecewise approximation to provide (\ref{operate}): for a given $\omega\in U$, first find the $U_j$ s.t. $\omega\in U_j$, and then 
\begin{equation}
     \Psi_\theta(\omega)(\mathbf{x})=\psi^{(j)}(\mathbf{x},\omega)\approx
     \psi_{gs}^\omega(\mathbf{x}), \quad \omega\in U_j.\label{operator1}
\end{equation}

\subsection{A unified model with distillation}\label{sec:unified_model}
To store the operator for each phase, the piecewise trained model described above incurs significant storage costs. Specifically, the number of model parameters grows linearly with the number of phases. To address this limitation, we train a unified model distilled from the piecewise trained DNN, improving prediction efficiency and reducing storage demands.

\begin{figure}[t!]
	 \centering 
	\begin{minipage}{0.3\linewidth}
		\vspace{1pt}
		\centerline{\includegraphics[width=\textwidth]{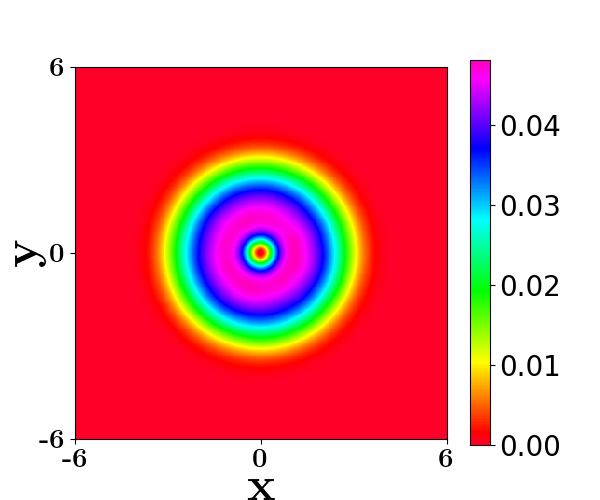}}
          
		\centerline{$\Omega=0.51$}
	\end{minipage}
	\begin{minipage}{0.3\linewidth}
		\vspace{1pt}
		\centerline{\includegraphics[width=\textwidth]{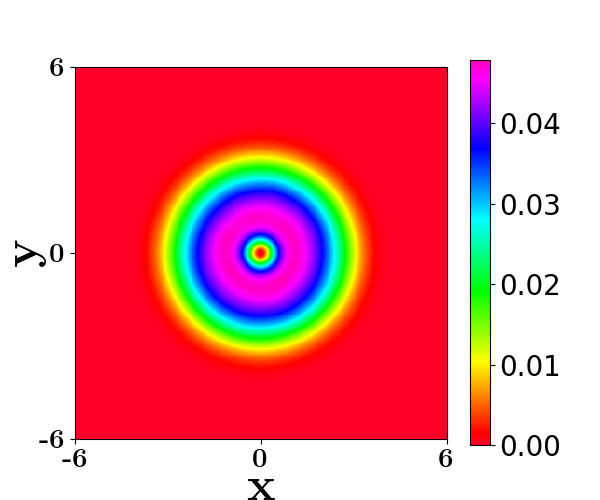}}
	 
		\centerline{$\Omega=0.5275$}
	\end{minipage}
        \begin{minipage}{0.3\linewidth}
		\vspace{1pt}
		\centerline{\includegraphics[width=\textwidth]{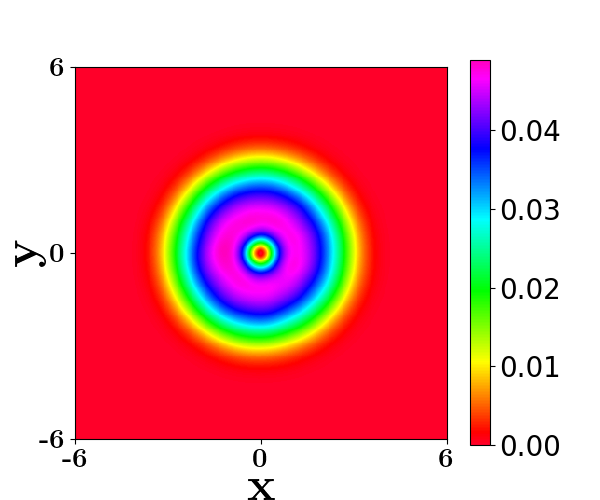}}
          
		\centerline{$\Omega=0.56  $}
	\end{minipage}

	\begin{minipage}{0.3\linewidth}
		\vspace{1pt}
		\centerline{\includegraphics[width=\textwidth]{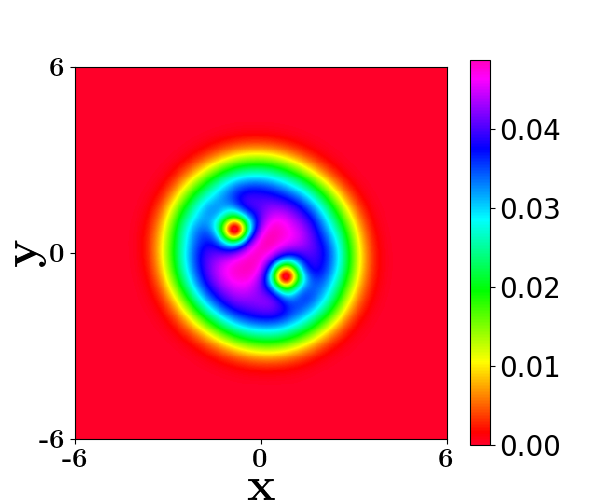}}
	 
		\centerline{$\Omega=0.585$}
	\end{minipage}
        \begin{minipage}{0.3\linewidth}
		\vspace{1pt}
		\centerline{\includegraphics[width=\textwidth]{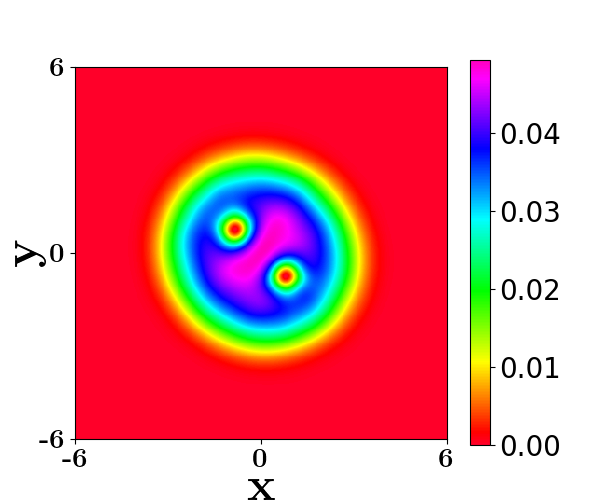}}
	 
		\centerline{$\Omega=0.5975$}
	\end{minipage}
        \begin{minipage}{0.3\linewidth}
		\vspace{1pt}
		\centerline{\includegraphics[width=\textwidth]{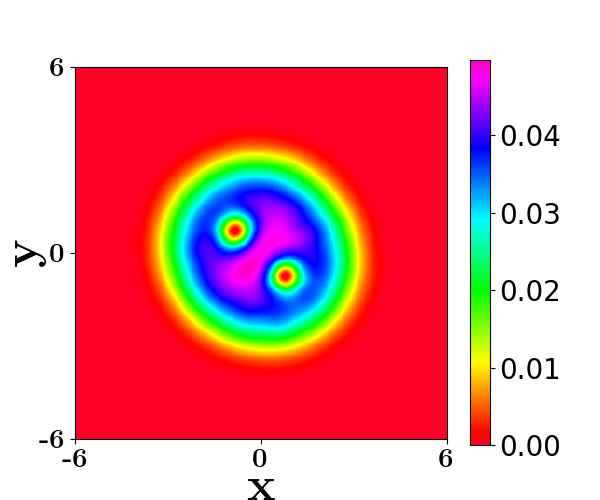}}
          
		\centerline{$\Omega=0.62$}
	\end{minipage}

	\begin{minipage}{0.3\linewidth}
		\vspace{1pt}
		\centerline{\includegraphics[width=\textwidth]{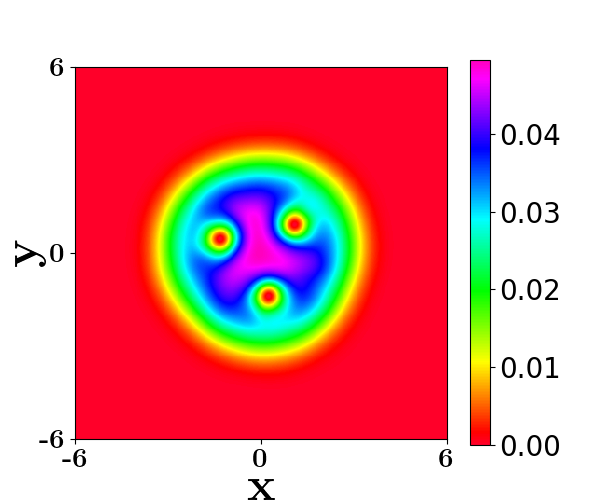}}
	 
		\centerline{$\Omega=0.63$}
	\end{minipage}
        \begin{minipage}{0.3\linewidth}
		\vspace{1pt}
		\centerline{\includegraphics[width=\textwidth]{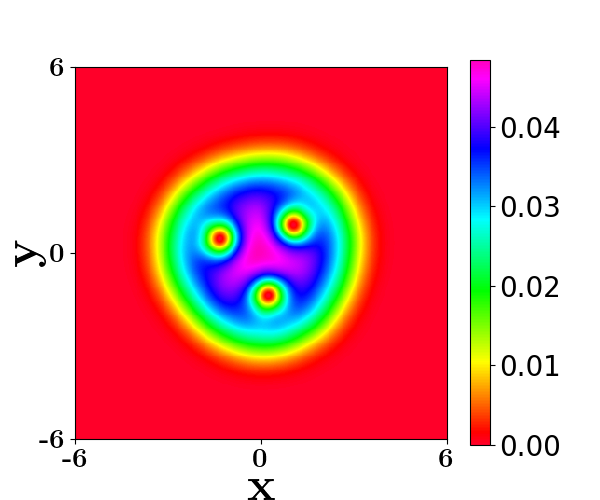}}
	 
		\centerline{$\Omega=0.6475$}
	\end{minipage}
        \begin{minipage}{0.3\linewidth}
		\vspace{1pt}
		\centerline{\includegraphics[width=\textwidth]{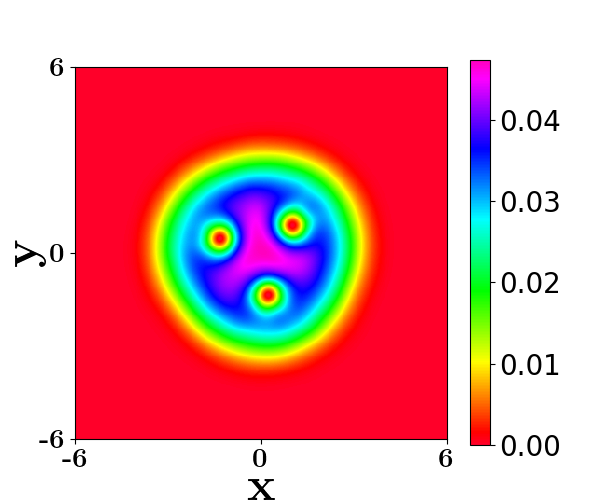}}
          
		\centerline{$\Omega=0.67$}
	\end{minipage}

	\caption{Surface plot of the distillation solution $|\psi_{\eta}|^2$ for different values of $\Omega$.}
	\label{fig_omega}
\end{figure}

\subsubsection{Distillation}\label{Distillation}
To compress $\{\psi^{(j)}\}_{j}$ into a smaller one, we first sample $N_{\omega}$ distinct parameters $\{\omega_i\}_{i=1}^{N_{\omega}}\subset \bigcup_j U_j$. Using the piecewise trained model with \eqref{operator1}, we directly generate a labeled dataset $\{\omega_i, \Psi_{\theta}(\omega_i)(\mathbf{x}),\nabla_{\mathbf{x}}\Psi_{\theta}(\omega_i)(\mathbf{x})\}_{i=1}^{N_{\omega}}$. Subsequently, a unified model $\psi_{\eta}(\mathbf{x},\omega)$ can be obtained by minimizing the following regression loss as in the Sobolev training~\cite{czarnecki2017sobolev}:
\begin{equation}\label{distillation}
    \mathcal{L}_{dist}(\eta):=\sum_{i=1}^{N_{\omega}}(\Vert\psi_{\eta}(\mathbf{x},\omega_i)- \Psi_{\theta}(\omega_i)(\mathbf{x})\Vert_{l^2}^2 +\Vert \nabla_{\mathbf{x}}\psi_{\eta}(\mathbf{x},\omega_i)- \nabla_{\mathbf{x}}\Psi_{\theta}(\omega_i)(\mathbf{x})\Vert_{l^2}^2).
\end{equation}

Through this model distillation process, the knowledge encapsulated in the original piecewise trained model (the teacher model) will be effectively transferred to a unified model (the student model), maintaining acceptable predictive accuracy while substantially reducing model complexity. The following two examples demonstrate its effectiveness. The network architecture here is consistent with that described in Section \ref{neural network}, and the initial learning rate is set to 0.001 with a decay rate of 0.98.

\begin{remark}
Among the labeled data for distillation (\ref{distillation}), 
the generalization capability of each $\psi^{(j)}$ within its corresponding domain $U_j$ allows to produce states $\Psi_\theta(\omega)(\mathbf{x})$ with $\omega$ outside of the original training set, i.e. the number of samples $N_{\omega}$ in \eqref{distillation} can be much larger than $M$ in \eqref{loss3}. Our approach as a whole is entirely unsupervised, in contrast to the data preparation process described in \cite{bai2025rapid}, which relies on labeled data obtained by solving \eqref{GS def} using the traditional imaginary-time evolution method—a computationally expensive procedure in the rotating frame.
\end{remark}

\begin{figure}[t!]
	 \centering 
	\begin{minipage}{0.3\linewidth}
		\vspace{1pt}
		\centerline{\includegraphics[width=\textwidth]{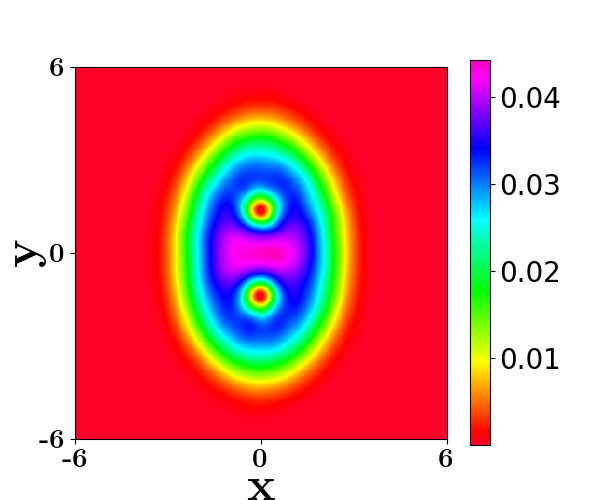}}
          
		\centerline{$\gamma_y=0.6$}
	\end{minipage}
	\begin{minipage}{0.3\linewidth}
		\vspace{1pt}
		\centerline{\includegraphics[width=\textwidth]{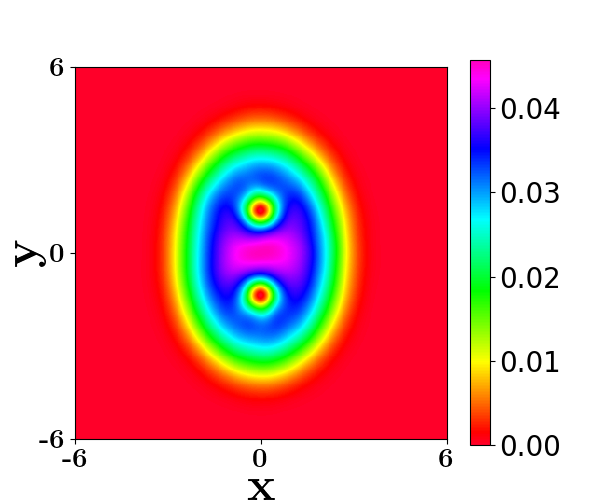}}
	 
		\centerline{$\gamma_y=0.65$}
	\end{minipage}
        \begin{minipage}{0.3\linewidth}
		\vspace{1pt}
		\centerline{\includegraphics[width=\textwidth]{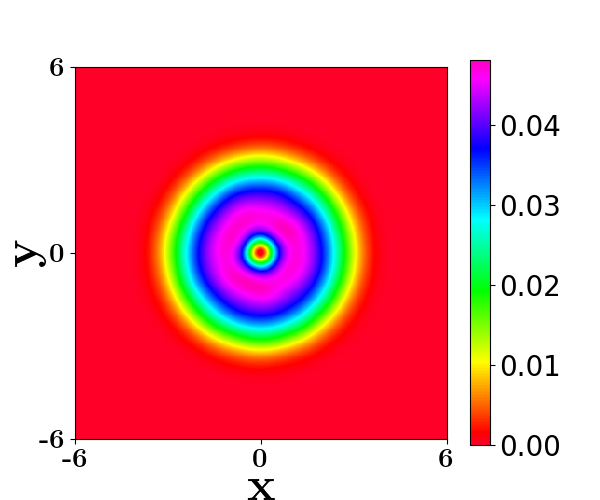}}
          
		\centerline{$\gamma_y=0.95$}
	\end{minipage}

        \begin{minipage}{0.3\linewidth}
		\vspace{1pt}
		\centerline{\includegraphics[width=\textwidth]{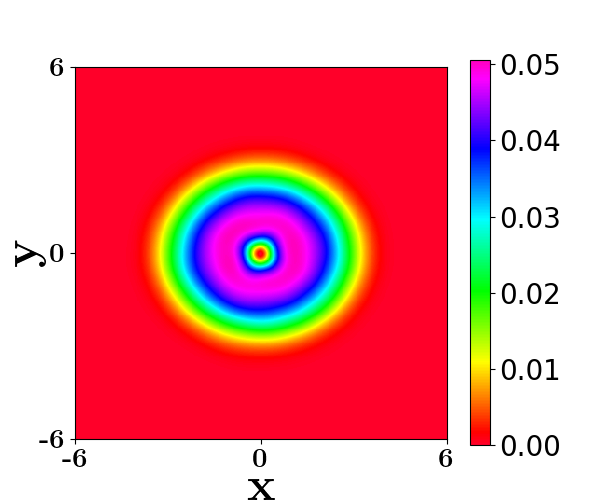}}
	 
		\centerline{$\gamma_y=1.25$}
	\end{minipage}
        \begin{minipage}{0.3\linewidth}
		\vspace{1pt}
		\centerline{\includegraphics[width=\textwidth]{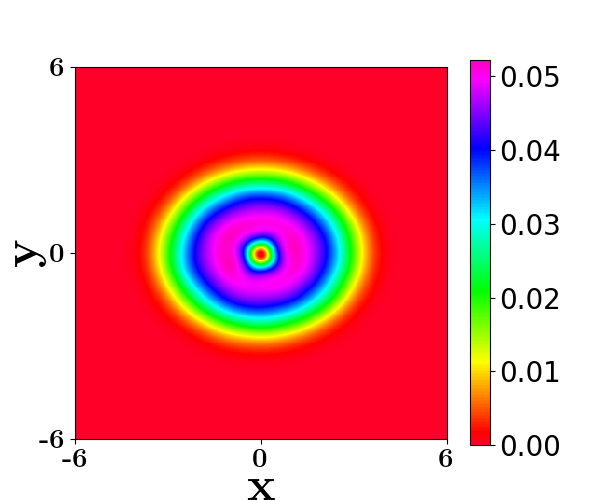}}
          
		\centerline{$\gamma_y=1.35$}
	\end{minipage}
        \begin{minipage}{0.3\linewidth}
		\vspace{1pt}
		\centerline{\includegraphics[width=\textwidth]{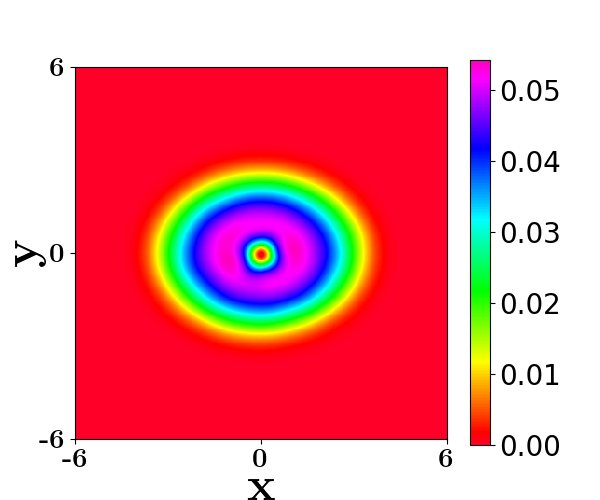}}
          
		\centerline{$\gamma_y=1.5$}
	\end{minipage}

	\caption{Surface plot of distillation solution $|\psi_{\eta}|^2$ for different values of $\gamma_y$.}
	\label{fig_gamma}
\end{figure}

\begin{exmp}[Generalizing $\Omega$ in multi-phase]
Consider the generalization of $\Omega$ for the 2D GS in Example \ref{exa1}. We take $\gamma_y=1$ and $U_1=[0.5,0.575]$, $U_2=[0.58,0.625]$, $U_3=[0.63,0.68]$. To construct the dataset by \eqref{operator1} for (\ref{distillation}), we choose $\{\Omega \in \bigcup_j^3 U_j |\Omega = 0.5+j*0.005, j\in \mathbb{N}\}$ as the parameter set. The prediction result from the distillation model $\psi_\eta$ is shown in Fig. \ref{fig_omega} for testing a series of $\Omega\in\bigcup_j^3 U_j$. \label{omega_gene}
\end{exmp}

\begin{exmp}[Generalizing $\gamma_y$ in multi-phase]
Consider the generalization of $\gamma_y$ for the 2D GS in Example \ref{exa2}. We take $\Omega=0.5$ and $U_1=[0.6, 0.7]$, $U_2=[0.8,1.6]$. The dataset for (\ref{distillation}) is $\{\gamma_{y} \in \bigcup_j^2 U_j |\gamma_y = 0.6+j*0.25, j\in \mathbb{N}\}$. The prediction result from $\psi_\eta$ is shown in Fig.
\ref{fig_gamma}.
\label{gamma_gene}
\end{exmp}

\begin{table}[h!]
  \begin{center}
    \begin{tabular}{|c|c|c|c|c|} 
    \hline
      Rotation &   \multicolumn{2}{c|}{$\Omega=0.9$} &  \multicolumn{2}{c|}{$\Omega=0.65$} \\
      \hline
      Method & PCG & unified DNN model & PCG & unified DNN model \\
      \hline
      Time(s) & $12.9$ & $0.004$  & $7.7$ & $0.004$  \\
      \hline
      Error & $4.6E{-3}$ & $4.6E{-3}$  & $1.3E{-3}$  & $1.2E{-3}$  \\
      \hline
    \end{tabular}
    \caption{Efficiency illustration: the error (\ref{energy err}) and the time of computation (inference for DNN).}
    \label{table_method}
  \end{center}
\end{table}

According to Fig.~\ref{fig_omega} and Fig.~\ref{fig_gamma}, the unified model predicts the right phase structure of GS. Specifically, for each $\omega_i$, the numerical solution $\psi_{\eta}$ shows the correct number of vortices corresponding to $U_i$. Notably, the numerical result from $\psi_{\eta}$ exhibits small localized irregularity in the figures, reflecting the existence of generalization error which will be further refined in the next part. 
Nevertheless, the unified operator network paves a way for fast and robust predictions of GS. To highlight its efficiency,  Table \ref{table_method} presents the computational time and accuracy of our trained model and the PCG method \cite{PCG} on Example \ref{omega_gene}. (The result of $\Omega=0.9$ for DNN was obtained via distillation analogously and compared with PCG on the same $64\times64$ uniform grid). 

\subsubsection{Distilled model as foundation model}\label{warm_start}
The above distillation process yields a lightweight and unified model $\psi_{\eta}$, which can be readily applied to GS prediction when high precision is not needed, e.g., the seek for only the vortex phase.

To further improve the generalization error, we can regard the distilled model as a pre-trained foundation model and employ fine-tuning strategy to further enhance prediction accuracy. Specifically, for a given physical parameter $\omega_0$, we first initialize $\psi_{\eta}(\mathbf{x}, \omega_0)$ using the network parameters $\eta^*$ obtained from the distillation process in \eqref{distillation}. We then train this model by minimizing the loss function in \eqref{loss1} with $\omega = \omega_0$. This procedure is summarized in Algorithm \ref{power1}.

\begin{algorithm}[!ht]
    \renewcommand{\algorithmicrequire}{\textbf{Input:}}
	\renewcommand{\algorithmicensure}{\textbf{Output:}}
	\caption{Fine-tune the distilled model for a given physical parameter}
    \label{power1}
    \begin{algorithmic}[1] 
        \REQUIRE Given $\omega$ and domain $G$; a distilled model $\psi_{\eta^*}$. 
	    \ENSURE A refined model $\psi_{\eta}\approx\psi_{gs}^{\omega}$. 
        
        \STATE Sample points for training: $\{\mathbf{x_i}\}_{i=1}^N\subset G$;
        \STATE Initialize $\psi_{\eta}=\psi_{\eta^*}$;
        \STATE Fine-tune $\psi_{\eta}$ by minimizing the loss in \eqref{loss1}.
    \end{algorithmic}
\end{algorithm}

\begin{figure}[htpb]
    \centering
    \includegraphics[width=0.9\linewidth]{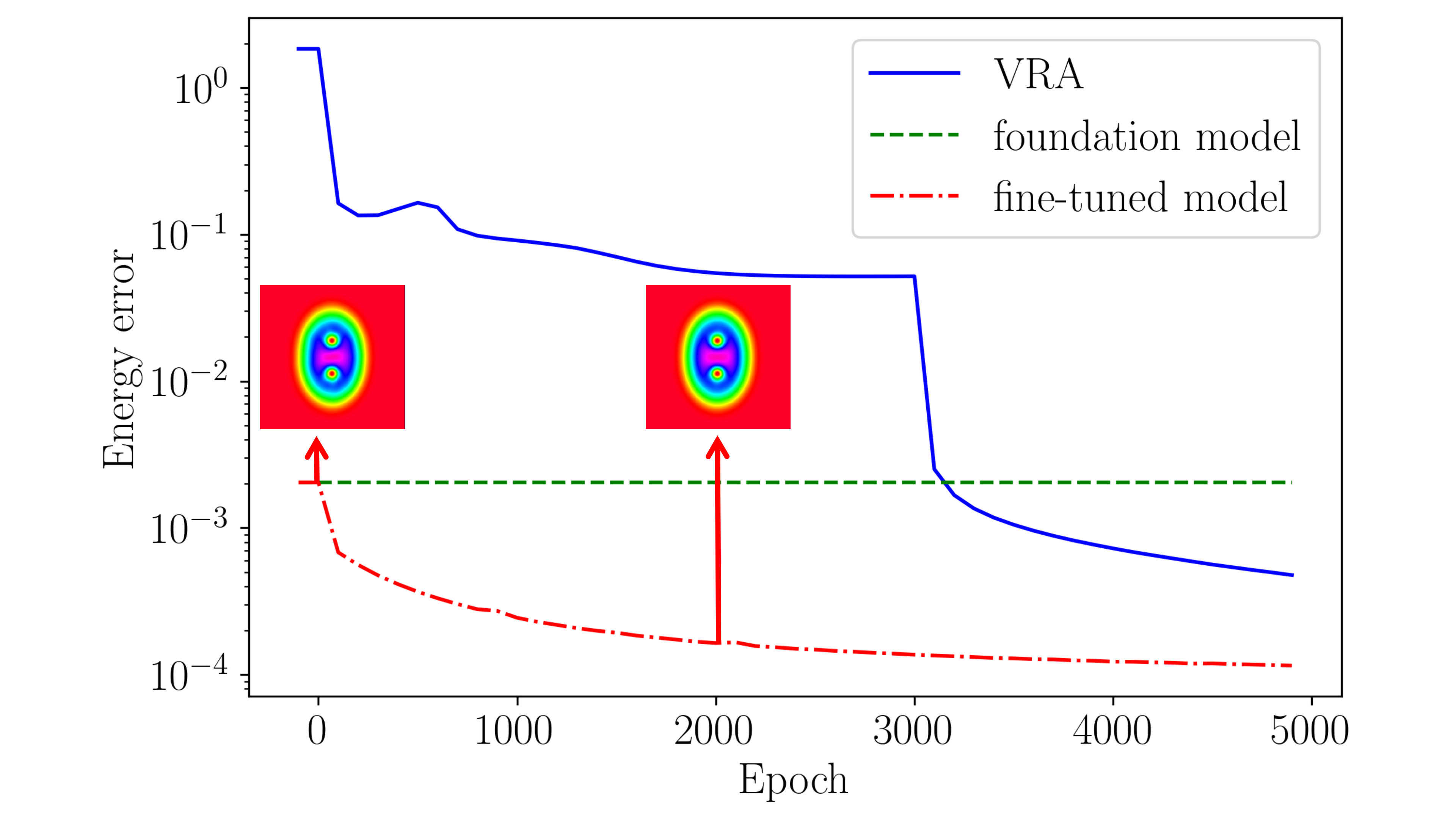}
    \caption{The energy error (\ref{energy err}) of different methods: the VRA method, the foundation model, and the fine-tuned model.}
    \label{erros_gene}
\end{figure}

For demonstration, we revisit the 2D GS problem with $\Omega = 0.5$ and $\gamma_y = 0.65$. The model $\psi_{\eta}$ is initialized using the distilled model $\psi_{\eta^*}$ from Example~\ref{gamma_gene}, and is further trained for 5,000 epochs using the loss function \eqref{loss1}. Fig.~\ref{erros_gene} illustrates the energy error curve of $\psi_{\eta}$ during training (labeled `fine-tuned model'). For comparison, we include the results of the VRA method from Section~\ref{subsection:VRA} (labeled `VRA') and the energy of the foundation model $\psi_{\eta^*}$ (labeled `foundation model').

As shown, the fine-tuned model starts with an error of $10^{-2}$ (from the foundation model) and rapidly converges to a significantly lower error of $10^{-4}$ after just 2,000 fine-tuning steps. In contrast, the VRA method initially reduces the error to $5 \times 10^{-2}$ during the Virtual stage, followed by further reduction in the Pullback stage. 


 Overall, the unified model provides rapid and valid prediction of GS, which can act as a good initial guess leading to more efficient and accurate further computation of GS. 

\subsection{Application to inverse problems} \label{inverse problem}    
Last but not least, we consider to solve the parameter inversion problem within BEC. 
Traditional numerical methods for inverse problems are frequently constrained by challenges associated with obtaining the forward operator, and the unified model established in Section \ref{sec:unified_model} becomes a natural candidate here.

We formalize the inverse problem as follows: under the conditions of known energy or image contour of the wave function (i.e., $\{\vert\psi_{gs}(\mathbf{x}_i)\vert\}_{i=1}^M$), we aim to reconstruct the physical parameters. 
For illustration, let us consider here the recovery of the trapping frequency $\gamma_y$ with other physical setup given and fixed.
We use $\mathcal{E}(\gamma)$ to represent the energy of the GS solution (\ref{GS def}) with $\gamma_y=\gamma$, and we can present its monotonicity as a function.

\begin{theorem}
Consider the 2D GS problem with a fixed $\Omega$ and varying $\gamma$. Let $D \subset \mathbb{R}_{+} \cup {0}$ be an interval on which $\mathcal{E}(\gamma)$ is defined. Then, $\mathcal{E}(\gamma)$ is strictly monotonically increasing and concave over $D$.\label{lemma}
\end{theorem}

\begin{proof}
    We first prove the strict monotonicity, i.e., $\forall \gamma < \gamma_1$, where ${\gamma}, {\gamma}_1\in D$, $\mathcal{E}({\gamma}) < \mathcal{E}(\gamma_1)$. By definition, $\mathcal{E}({\gamma})$ is obtained as
       $\mathcal{E}(\gamma)=E_{\gamma}(\psi^{\gamma}_{gs}),$
    where
       $$ E_{\gamma}(\psi)=\int_{\mathbb{R}^{2}} \left[ \frac{1}{2}|\nabla\psi|^{2} + \frac{x_1^2+\gamma x_2^2}{2}|\psi|^{2} + \frac{\beta}{2}|\psi|^{4} - \Omega \overline{\psi} L_{z} \psi \right] d\mathbf{x},$$
    and 
$\psi^{\gamma}_{gs}:=\mathop{\arg\min}\limits_{\Vert\psi\Vert_2=1}E_\gamma(\psi)$. Consequently, we have 
    \begin{equation}
        E_{\gamma}(\psi^{\gamma}_{gs})\leq E_{\gamma}(\psi^{\gamma_1}_{gs}).\label{mono_1}
    \end{equation}
     Besides, since
    \begin{equation}
        E_\gamma(\psi)=\int_{\mathbb{R}^{2}} \left[ \frac{1}{2}|\nabla\psi|^{2} + \frac{x_1^2}{2}|\psi|^{2} + \frac{\beta}{2}|\psi|^{4} - \Omega \overline{\psi} L_{z} \psi \right]d\mathbf{x}+\gamma\int_{\mathbb{R}^{d}}\frac{x_2^2}{2}\vert\psi\vert^2 d\mathbf{x}\label{linear}
    \end{equation}
    and $\int_{\mathbb{R}^{2}}\frac{x_2^2}{2}\vert\psi\vert^2 d\mathbf{x}$ is positive, we have 
    \begin{equation}
        E_{\gamma}(\psi^{\gamma_1}_{gs}) < E_{\gamma_1}(\psi^{\gamma_1}_{gs}).\label{mono_2}
    \end{equation}
    By combining \eqref{mono_1} and \eqref{mono_2}, we obtain 
    
       $$ \mathcal{E}(\gamma)=E_{\gamma}(\psi^{\gamma}_{gs})\leq E_{\gamma}(\psi^{\gamma_1}_{gs})\leq E_{\gamma_1}(\psi^{\gamma_1}_{gs})=\mathcal{E}(\gamma_1).$$

    We now proceed to prove the concavity. According to the equivalent definition of concave functions, we only need to prove that: $\forall b\in \mathcal{D}$, there is a linear function $ \mathcal{G}(\gamma)$ s.t. $\mathcal{G}(b) = \mathcal{E}(b)$ and $\mathcal{E}(\gamma)\leq \mathcal{G}(\gamma), \forall \gamma\in D$. In fact, by defining 
      $  \mathcal{G}(\gamma) =E_\gamma(\psi_{gs}^b)$ and using \eqref{linear}, we have that $\mathcal{G}(\gamma)$ is a linear function of $\gamma$ with $\mathcal{G}(b) = \mathcal{E}(b)$. Moreover, $\mathcal{E}(\gamma)=E_\gamma(\psi_{gs}^{\gamma})\leq E_{\gamma}(\psi_{gs}^b)=\mathcal{G}(\gamma), \forall \gamma\in D$. This verifies the concavity.
\end{proof}

The continuity and strict monotonicity of $\mathcal{E}(\gamma)$ indicates its unique solvability for $\gamma_y$ with a given energy value. Traditional approach like the bisection method can be employed to solve such a nonlinear equation for $\gamma$, as long as the expression of $\mathcal{E}(\gamma)$ is explicitly known which is approximated here by  the unified model $\psi_{\eta}$, i.e., 
$$\mathcal{E}(\gamma)\approx \mathcal{E}_{\eta}(\gamma) := E_{\gamma}(\psi_{\eta}(\mathbf{x},\gamma)).$$
Denoting $\mathcal{E}_{obs}$ the observed/given energy data, the nonlinear equation to solve reads
\begin{equation}\label{eq:e_eta_E}
    \mathcal{E}_{\eta}(\gamma) = \mathcal{E}_{obs}.
\end{equation}

\begin{exmp}[Inversion of $\gamma_y$ with observed energy] Consider a 2D GS problem with $\Omega = 0.5,\,\beta=100$ and $\gamma=\gamma_y$ unknown. Given the observed GS energy $\mathcal{E}_{obs}=4.3243$ and $\Omega = 0.5$, we recover $\gamma$ by solving \eqref{eq:e_eta_E}. As a reference,  $\mathcal{E}_{obs}$ corresponds to $\gamma = 1.3$.
\end{exmp}

With the initial interval set to $[0.8, 1.6]$ and chosen the unified model $\psi_{\eta}$ from Example~\ref{gamma_gene}, the bisection method after merely 10 iterations obtains $\gamma$ as $1.2996$, achieving an error at the order $10^{-4}$.

\begin{exmp}[Inverse problems with known image] Under the same setup of the above example, but without the data of energy, we are given with a $64\times64$ image matrix of GS, i.e., the values of $|\psi_{gs}|$ on grids. We still look for the parameter $\gamma$.\label{inverse noise}
\end{exmp}

Note that now the functional value of $|\psi_{gs}|$ is not enough to give the value of energy (\ref{linear}). What can be deduced by (\ref{linear}) is 
$\mathcal{E}^{\prime}(\gamma)=\int_{\mathbb{R}^{d}}\frac{x_2^2}{2}\vert\psi\vert^2 d\mathbf{x}.$
Then using a finite difference approximation together with the help of the unified model, we consider $$ \mathcal{E}^{\prime}(\gamma)\approx \frac{1}{\delta\gamma}\left(\mathcal{E}_\eta(\gamma+\delta\gamma)-\mathcal{E}_\eta(\gamma)\right).$$ 
Starting from the initial interval $[0.8,1.6]$ and $\delta \gamma=0.8/3$, after 10 refinements, the  result of $\gamma_1$ is found  within the interval $[1.2975, 1.3025]$, with an error of order $10^{-3}$. 

  

\section{Conclusion} \label{conclusion}
The paper considers the computation for the ground state (GS) of rotating Bose-Einstein condensates (BEC) via unsupervised deep learning. The mass constraint in the energy minimization problem and the presence of rotation that induces quantized vortices in GS with phase transitions sensitive to the rotating speed, challenge the accuracy and efficiency of computation. To enable accurate learning with robustness across phases, we introduced a normalized loss function to enforce the mass constraint precisely, and a vital training strategy named virtual rotation acceleration to guide the training towards the correct vortex phase. 
Numerical experiments in various physical conditions demonstrated the effectiveness and accuracy of the proposed approach, and comparisons were made with the normalized network method from the literature. 
Eventually, through distillation, a unified operator network has been established to generalize physical parameters, which is capable of rapidly predicting GS across different phases. The applicability of the model in inverse problems, specifically parameter identification based on observed energy or solution images, has been shown in the end. 

\bibliographystyle{siamplain}
\bibliography{references}
\end{document}